\documentclass[a4paper,UKenglish]{lipics}
\usepackage{microtype}%if unwanted, comment out or use option "draft"

\usepackage{dsfont}

\usepackage{mathpartir}
\usepackage{stmaryrd}
\usepackage{amssymb}
\usepackage{amsmath}

\usepackage{listings}
\usepackage{booktabs}

\usepackage{paralist}

\usepackage{tikz}
\usetikzlibrary{shapes,arrows,fit,backgrounds}

\usepackage[colorinlistoftodos]{todonotes}

\usepackage{url}

\usepackage{soul}

\usepackage{color, colortbl}

\definecolor{Light}{gray}{.90}
\sethlcolor{Light}

\newcommand{\Label}[1]{\textit{#1}}

\newcommand{\TJS}{\textsf{TreatJS}}
\newcommand{\SBX}{\textsf{DecentJS}}

\lstdefinelanguage{JavaScript}{
  keywords={
    attributes, class, classend, do, empty, endif, endwhile, fail,
    function, functionend, if, implements, in, inherit, inout, not, of,
    operations, out, return, then, types, while, use, else, switch, case,
    break, default, for, var, with, new, eval},
  keywordstyle=\color{black}\bfseries,
  ndkeywords={
    use, strict
  },
  ndkeywordstyle=\color{black}\bfseries,
  identifierstyle=\color{black},
  sensitive=true,
  comment=[l]{//},
  morecomment = [s]{/*}{*/},
  morecomment = [s][\color{gray}]{/**}{*/},
  commentstyle=\color{gray},%\ttfamily,
  stringstyle=\color{black},%\ttfamily
  basicstyle=\itshape,
}
\lstdefinelanguage{HTML5}{
  language=html,
  morecomment=[s]{<!--}{-->},
  escapeinside={*'}{'*},
}

\newcolumntype{I}{>{\iffalse}c<{\fi}@{~}}
\newcolumntype{H}{>{\setbox0=\hbox\bgroup}c<{\egroup}@{}}

\newcommand\bbc{::=}

\newcommand{\eval}{~\Downarrow~} %% conflict with semantic package
\newcommand{\entails}{~\vdash~}

\newcommand{\dom}[1]{\textit{dom(}#1\textit{)}}

\newcommand{\lj}{\lambda_{J}}
\newcommand{\lsbx}{\lambda^{\textit{SBX}}_{J}}

\newcommand{\ljExpe}{e}
\newcommand{\ljExpf}{f}
\newcommand{\ljExpg}{g}
\newcommand{\ljExp}{\ljExpe}

\newcommand{\ljConst}{c}

\newcommand{\ljVarx}{x}
\newcommand{\ljVary}{y}
\newcommand{\ljVar}{\ljVarx}

\newcommand{\ljOp}[2]{\textit{op}(#1,#2)}
\newcommand{\ljAbs}[2]{\lambda#1.#2}
\newcommand{\ljApp}[2]{#1(#2)}
\newcommand{\ljNew}[1]{\textit{new}~#1}
\newcommand{\ljGet}[2]{#1[#2]}
\newcommand{\ljPut}[3]{#1[#2]=#3}

\newcommand{\ljWrap}[1]{\textit{wrap}(#1)}
\newcommand{\ljFresh}[1]{\textit{fresh}~#1}
\newcommand{\ljRecomp}[1]{\textit{compile}(#1)}

\newcommand{\ljSbx}[2]{\Lambda#1.#2}

\newcommand{\ljValu}{u}
\newcommand{\ljValv}{v}
\newcommand{\ljValw}{w}
\newcommand{\ljVal}{\ljValv}

\newcommand{\ljSValu}{\widehat\ljValu}
\newcommand{\ljSValv}{\widehat\ljValv}
\newcommand{\ljSValw}{\widehat\ljValw}
\newcommand{\ljSVal}{\ljSValv}

\newcommand{\ljLoc}{l}
\newcommand{\ljSLoc}{\widehat\ljLoc}

\newcommand{\ljClosure}{f}
\newcommand\ljNoClosure{-}

\newcommand{\ljDic}{d}

\newcommand{\ljObj}{o}

\newcommand{\ljTerm}{t}

\newcommand{\sbx}{\mathcal{S}}

\newcommand{\ljUndefined}{\textit{undefined}}
\newcommand{\ljNull}{\textit{null}}

\newcommand{\ljStore}{\sigma}

\newcommand{\ljEnv}{\rho}
\newcommand{\ljSEnv}{\widehat{\ljEnv}}

\newcommand{\eqenv}[2]{#1\simeq_{\ljEnv}#2}
\newcommand{\eqstore}[2]{#1\simeq#2}

\newcommand{\RuleLjConst}{Const}

\newcommand{\RuleLjVar}{Var}

\newcommand{\RuleLjOp}{Op}
\newcommand{\RuleLjOpE}{Op-E}
\newcommand{\RuleLjOpF}{Op-F}

\newcommand{\RuleLjNewE}{New-E}
\newcommand{\RuleLjNew}{New}

\newcommand{\RuleLjAbs}{Abs}

\newcommand{\RuleLjAppE}{App-E}
\newcommand{\RuleLjAppF}{App-F}
\newcommand{\RuleLjApp}{App}

\newcommand{\RuleLjGetE}{Get-E}
\newcommand{\RuleLjGetF}{Get-F}
\newcommand{\RuleLjGetProto}{Get-Proto}
\newcommand{\RuleLjGetUndef}{Get-Undef}
\newcommand{\RuleLjGet}{Get}

\newcommand{\RuleLjPutE}{Put-E}
\newcommand{\RuleLjPutF}{Put-F}
\newcommand{\RuleLjPutG}{Put-G}
\newcommand{\RuleLjPut}{Put}

\newcommand{\RuleSbxFreshE}{Sandbox-Fresh-E}
\newcommand{\RuleSbxFresh}{Sandbox-Fresh}
\newcommand{\RuleSbxAbs}{Sandbox-Abstraction}
\newcommand{\RuleSbxAbb}{Sandbox-Application}

\newcommand{\RuleWrapConst}{Wrap-Const}
\newcommand{\RuleWrapSandbox}{Wrap-Sandbox}
\newcommand{\RuleWrapNonProxy}{Wrap-NonProxyObject}
\newcommand{\RuleWrapProxy}{Wrap-ProxyObject}
\newcommand{\RuleWrapExisting}{Wrap-Existing}

\newcommand{\RuleRecompNonFunction}{Recompile-NonFunctionObject}
\newcommand{\RuleRecompFunction}{Recompile-FunctionObject}
\newcommand{\RuleRecompileProxy}{Recompile-ProxyObject}
\newcommand{\RuleRecompileExisting}{Recompile-Existing}

\newcommand{\RuleAppSandbox}{App-Sandbox}
\newcommand{\RuleGetShadow}{Get-Shadow}
\newcommand{\RuleGetSandbox}{Get-Sandbox}
\newcommand{\RulePutSandbox}{Put-Sandbox}

\title{Transaction-based Sandboxing for JavaScript\newline\textnormal{Technical Report}
}
\titlerunning{Transaction-based Sandboxing for JavaScript} 

\author{Matthias Keil}
\author{Peter Thiemann}
\affil{University of Freiburg\\Freiburg, Germany\\
  \texttt{\{keilr,thiemann\}@informatik.uni-freiburg.de}
}

\authorrunning{M. Keil and P. Thiemann}

\Copyright{Matthias Keil and Peter Thiemann}

\begin{document}

\maketitle

%          _         _                  _   
%    /\   | |       | |                | |  
%   /  \  | |__  ___| |_ _ __ __ _  ___| |_ 
%  / /\ \ | '_ \/ __| __| '__/ _` |/ __| __|
% / ____ \| |_) \__ \ |_| | | (_| | (__| |_ 
%/_/    \_\_.__/|___/\__|_|  \__,_|\___|\__|

\begin{abstract}
  Today's JavaScript applications are composed of scripts from different origins
  that are loaded at run time. As not all of these origins are equally trusted,
  the execution of these scripts should be isolated from one another. However,
  some scripts must access the application state and some may be allowed to
  change it, while preserving the confidentiality and integrity constraints of
  the application.

  This paper presents design and implementation of \SBX, a language-embedded
  sandbox for full JavaScript. It enables scripts to run in a configurable
  degree of isolation with fine-grained access control. It provides a
  transactional scope in which effects are logged for review by the access
  control policy. After inspection of the log, effects can be committed to the
  application state or rolled back.

  The implementation relies on JavaScript proxies to guarantee full
  interposition for the full language and for all code, including dynamically
  loaded scripts and code injected via \lstinline{eval}. Its only restriction is
  that scripts must be compliant with JavaScript's strict mode.
\end{abstract}

% _____       _                 _            _   _             
%|_   _|     | |               | |          | | (_)            
%  | |  _ __ | |_ _ __ ___   __| |_   _  ___| |_ _  ___  _ __  
%  | | | '_ \| __| '__/ _ \ / _` | | | |/ __| __| |/ _ \| '_ \ 
% _| |_| | | | |_| | | (_) | (_| | |_| | (__| |_| | (_) | | | |
%|_____|_| |_|\__|_|  \___/ \__,_|\__,_|\___|\__|_|\___/|_| |_|

\section{Introduction}
\label{sec:introduction}

JavaScript is used by 93.1\%\footnote{according to http://w3techs.com/, status
as of March 2016} of all the websites. Most of them rely on third-party
libraries for connecting to social networks, feature extensions, or
advertisement. Some of these libraries are packaged with the application, but
others are loaded at run time from origins of different trustworthiness,
sometimes depending on user input. To compensate for different levels of trust,
the execution of dynamically loaded code should be isolated from the application
state.

Today's state of the art in securing JavaScript applications that include code
from different origins is an all-or-nothing choice. Browsers apply protection
mechanisms, such as the same-origin policy~\cite{SameOriginPolicy} or the signed
script policy~\cite{SignedScriptsInMozilla}, so that scripts either run in
isolation or gain full access.

While script isolation guarantees noninterference with the working of the
application as well as preservation of data integrity and confidentiality, there
are scripts that must have access to part of the application state to function
meaningfully. As all included scripts run with the same authority, the
application script cannot exert fine-grained control over the use of data by an
included script.

Thus, managing untrusted JavaScript code has become one of the key challenges of
present research on JavaScript~\cite{AgtenVanAckerBrondsemaPhungDesmetPiessens2012,HedinBirgissonBelloSabelfeld2014,DewaldHolzFreiling2010,DhawanGanapathy2009,RichardsHammerNardelliJagannathan2013,MeyerovichLivshits2010,PhungSandsChudnov2009,MagaziniusPhungSands2010,MaffeisTaly2006,GuarnieriLivshits2009}. Existing approaches are either based on restricting JavaScript code to a
statically verifiable language subset (e.g., Facebook's FBJS~\cite{FaecbookJS}
or Yahoo's ADsafe~\cite{ADsafe}), on enforcing an execution model that only
forwards selected resources into an otherwise isolated compartment by  filtering
and rewriting like Google's Caja project~\cite{GoogleCaja}, or on
implementing monitoring facilities inside the JavaScript engine~\cite{RichardsHammerNardelliJagannathan2013}.

However, these approaches have known deficiencies: the first two need to restrict usage
of JavaScript's dynamic features, they do not apply to code generated at run time,
and they require extra maintenance efforts because their analysis and
transformation needs to be kept in sync with the evolution of the
language. Implementing monitoring in the JavaScript engine is fragile
and incomplete: while efficient, such a solution only works for one
engine and it is hard to maintain due to the high activity in engine
development and optimization.

\paragraph*{Contributions}
We present the design and implementation of \SBX, a sandbox for JavaScript
which enforces noninterference (integrity and confidentiality) by run-time
monitoring. Its design is inspired by revocable
references~\cite{CutsemMiller2010,Miller2006} and SpiderMonkey's compartment
concept~\cite{WagnerGalWimmerEichFranz2011}.

Compartments create a separate memory heap for each website, a technique initially
introduced to optimize garbage collection. All objects created by a website are
only allowed to touch objects in the same compartment. Proxies are the only
objects that can cross the compartment boundaries. They are used as cross
compartment wrappers to make objects accessible in other compartments.

\SBX\ adapts SpiderMonkey's compartment concept. Each sandbox implements a fresh
scope to run code in isolation to the application state. Proxies implement a
membrane~\cite{CutsemMiller2010,Miller2006} to guarantee full interposition and
to make objects accessible inside of a sandbox.

%  ___                  _            
% / _ \__ _____ _ ___ _(_)_____ __ __
%| (_) \ V / -_) '_\ V / / -_) V  V /
% \___/ \_/\___|_|  \_/|_\___|\_/\_/ 

\paragraph*{Outline of this Paper}

The paper is organized as follows:
Section~\ref{sec:primer} introduces \SBX's facilities from a programmer's point
of view.
Section~\ref{sec:sandbox} recalls proxies and membranes from related work and
explains the principles underlying the implementation.
Section~\ref{sec:limitations} discusses \SBX's
limitations and Section~\ref{sec:evaluation} reports our 
experiences from applying sandboxing to a set of
benchmark programs.
Finally, Section~\ref{sec:conclusion} concludes.

Appendix~\ref{sec:appendix/motivation} presents an example demonstrating the
sandbox hosting a third-party library. Appendix~\ref{sec:appendix/application}
shows some example scenarios that already use the implemented system.
Appendix~\ref{sec:appendix/semantics} shows the operational semantics of a core
calculus with sandboxing. Appendix~\ref{sec:appendix/technicalresults} states
some technical results.
Appendix~\ref{sec:relatedwork} discusses related work and
Appendix~\ref{sec:appendix/results} reports our experiences from applying 
sandboxing to a set of benchmark programs.

% _____      _                     
%|  __ \    (_)                    
%| |__) | __ _ _ __ ___   ___ _ __ 
%|  ___/ '__| | '_ ` _ \ / _ \ '__|
%| |   | |  | | | | | | |  __/ |   
%|_|   |_|  |_|_| |_| |_|\___|_|   
                                  
\section{Transaction-based Sandboxing: A Primer}
\label{sec:primer}

This section introduces transaction-based sandboxing and shows a series of
examples that explains how sandboxing works.

Transactional sandboxing is inspired by the idea of transaction processing in
database systems~\cite{WeikumVossen2001:_trans} and transactional
memory~\cite{ShavitTouitou1995}. Each sandbox implements a transactional scope
the content of which can be examined, committed, or rolled back.

Central to our sandbox is the implementation of a membrane on values that cross
the sandbox boundary. The membrane supplies effect monitoring and guarantees
noninterference. Moreover, it features identity preservation and handles shadow
objects. \emph{Shadow objects} allow sandbox-internal modifications of objects
without effecting there origins. The modified version is only visible inside of
the sandbox and different sandbox environments may manipulate the same object.

Sandboxing provides transactions, a unit of effects that represent the set of
modifications (write effects) on its membrane. Effects enable to check for
conflicts and differences, to rollback particular modifications, or to commit a
modification to its origin.

The implementation of the system is available on the
web\footnote{\url{https://github.com/keil/DecentJS}}.

% ___               _ _             _           
%/ __| __ _ _ _  __| | |__  _____ _(_)_ _  __ _ 
%\__ \/ _` | ' \/ _` | '_ \/ _ \ \ / | ' \/ _` |
%|___/\__,_|_||_\__,_|_.__/\___/_\_\_|_||_\__, |
%                                         |___/ 

\subsection{Cross-Sandbox Access}
\label{sec:primer-sandboxing}

\begin{figure}[t]
  \centering
  \begin{lstlisting}[name=primer]
  function Node (value, left, right) {
    this.value = value;
    this.left = left;
    this.right = right;
  }
  Node.prototype.toString = function () {
    return (this.left?this.left + ", ":"") + this.value +(this.right?", "+this.right:"");
  }
  function heightOf (node) {
    return Math.max(((node.left)?heightOf(node.left)+1:0), ((node.right)?heightOf(node.right)+1:0)); *'\label{line:nativecall}'*
  }
  function setValue (node) { *'\label{line:setvalue}'*
    if (node) {
      node.value=heightOf(node); *'\label{line:heightOf}'*
      setValue(node.left);
      setValue(node.right);
    }
  }
  \end{lstlisting}
  \caption{Implementation of \lstinline{Node}. Each node object consists of a
  value field, a left node, and a right node. Its prototype provides a
  \lstinline{toString} method that returns a string representation.
  Function \lstinline{heightOf} computes the height of a node and function
  \lstinline{setValue} replaces the value field of a node by its height,
  recursively.
}
  \label{fig:example}
\end{figure}

We consider operations on binary trees as defined by \lstinline{Node} in Figure \ref{fig:example} along with some auxiliary functions.
As an example, we perform operations on a tree consisting of one node
and two leaves. All value fields are initially \lstinline{0}.
\begin{lstlisting}[name=primer]
var root = new Node(0, new Node(0), new Node(0));*'\label{line:tree}'*
\end{lstlisting}

Next, we create a new empty sandbox by calling the constructor \lstinline|Sandbox|.
Its first parameter acts as the global object of
the sandbox environment. It is wrapped in a proxy to mediate all
accesses and it is placed on top of the scope chain for code executing
inside the sandbox. The second parameter is a configuration
object.  A sandbox is a first class value that can be used for several executions.

\begin{lstlisting}[name=primer]
var sbx = new Sandbox(this, {/* some parameters */});*'\label{line:freshsandbox}'*
\end{lstlisting}
One use of a sandbox is to wrap invocations of function objects. To
this end, the sandbox API provides methods \lstinline{call}, \lstinline{apply}, and
\lstinline{bind} analogous to methods from
\lstinline{Function.prototype}.
For example, we may call \lstinline{setValue} on \lstinline{root}
inside of  \mbox{\lstinline{sbx}.}
\begin{lstlisting}[name=primer]
sbx.call(setValue, this, root);*'\label{line:sandboxcall}'*
\end{lstlisting}
The first argument of \lstinline{call} is a function object that is decompiled and redefined inside
the sandbox. This step erases the function's free variable bindings and builds a 
new closure relative to the sandbox's global object. 
The second argument, the receiver object of the call, as well as the
actual arguments of the call are wrapped in proxies to make these objects accessible inside of the
sandbox.

The wrapper proxies mediate access to their target objects outside the sandbox.
Reads are forwarded to the target unless there are local
modifications. The return values are wrapped in proxies, again.
Writes produce a \emph{shadow value} (cf.
Section~\ref{sec:sandbox/shadow}) that represents the sandbox-internal modification of
an object. Initially, this modification is only visible to reads inside the sandbox.

Native objects, like the \lstinline{Math} object in line
\ref{line:nativecall}, are also wrapped in a proxy, but their methods cannot be decompiled because
there exists no string representation. Thus, native methods must either be trusted or forbidden. 
Fortunately, most native methods do not have side effects, so we chose
to trust them.

Given all the wrapping and sandboxing, the call in line
\ref{line:sandboxcall} did not modify the \lstinline{root} object:
\begin{lstlisting}[name=primer]
root.toString(); // returns 0, 0, 0
\end{lstlisting}
But calling \lstinline{toString} inside the sandbox shows the effect.
\begin{lstlisting}[name=primer]
sbx.call(root.toString, root); // return 0, 1, 0
\end{lstlisting}

% ___  __  __        _      
%| __|/ _|/ _|___ __| |_ ___
%| _||  _|  _/ -_) _|  _(_-<
%|___|_| |_| \___\__|\__/__/

\subsection{Effect Monitoring}
\label{sec:primer-effects}

During execution, each sandbox records the effects on objects that cross the sandbox
membrane.
The resulting lists of \emph{effect objects} are accessible through \lstinline{sbx.effects},
\lstinline{sbx.readeffects}, and \lstinline{sbx.writeeffects} which
contain all effects, read effects, and write effects, respectively. 
All three lists offer query methods to select the effects of
a particular object.
\begin{lstlisting}[name=primer]
sbx.call(heightOf, this, root); 
var rects = sbx.effectsOf(this);
print(";;; Effects of this");
rects.foreach(function(i, e) {print(e)});
\end{lstlisting}
The code snippet above prints a list of all effects performed on
\lstinline{this}, the global object, by executing the
\lstinline{heightOf} function on \lstinline{root}. The output shows
the resulting accesses to \lstinline{heightOf} and \lstinline{Math}.  
\begin{lstlisting}[name=primer]
;;; Effects of this
(1425301383541) has [name=heightOf]
(1425301383541) get [name=heightOf]
(1425301383543) has [name=Math]
(1425301383543) get [name=Math]
...
\end{lstlisting}
The first column shows a timestamp, the second shows the name of the effect,
and the last column shows the name of the requested parameter. The
list does not contain write accesses to \lstinline{this}.
But there are write effects to \lstinline{value} from the previous invocation of \lstinline{setValue}. 
\begin{lstlisting}[name=primer]
var wectso = sbx.writeeffectsOf(root);
print(";;; Write Effects of root");
wectso.foreach(function(i, e) {print(e)});
\end{lstlisting}

\begin{lstlisting}[name=primer]
;;; Write Effects of root
(1425301634992) set [name=value]  
\end{lstlisting}

%  ___ _                          
% / __| |_  __ _ _ _  __ _ ___ ___
%| (__| ' \/ _` | ' \/ _` / -_|_-<
% \___|_||_\__,_|_||_\__, \___/__/
%                    |___/        

\subsection{Inspecting a Sandbox}

The state inside and outside of a sandbox may diverge for different
reasons. We distinguish changes, differences, and
conflicts.

A \emph{change} indicates if the sandbox-internal value has been changed
with respect to the outside value. A \emph{difference}
indicates if the outside value has been modified after the sandbox has
concluded. 
For example, a difference to the previous execution of
\lstinline{setValue} arises if we replace the left leaf element by a
new subtree of height 1 outside of the sandbox.  
\begin{lstlisting}[name=primer]
root.left = new Node(new Node(0), new Node(0));*'\label{line:newtree}'* 
\end{lstlisting}

Changes and differences can be examined using an API that is very
similar to the effect API. There are flags to check whether a sandbox
has changes or differences as well as iterators over them.

A \emph{conflict} arises in the comparison between
different sandboxes. 
Two sandbox environments are in conflict if at least one sandbox modifies
a value that is accessed by the other sandbox later on.
We consider only Read-After-Write and Write-After-Write conflicts.
To demonstrate conflicts, we define a function
\lstinline{appendRight}, which adds a new subtree on the right.

\begin{lstlisting}[name=primer]
function appendRight (node) {
  node.right = Node('a', Node('b'), Node('c'));
}
\end{lstlisting}
To recapitulate, the global \lstinline{root} is still unmodified and prints
\lstinline{0,0,0,0,0}, whereas the \lstinline{root} in \lstinline{sbx} prints
\lstinline{0,0,0,1,0}. Now, let's execute \lstinline{appendRight} in a
new sandbox \lstinline{sbx2}.
\begin{lstlisting}[name=primer]
var sbx2 = new Sandbox(this, {/* some parameters */});
sbx2.call(appendRight, this, root);*'\label{line:sandboxcall2}'*
\end{lstlisting}
Calling \lstinline{toString} in \lstinline{sbx2}
prints \lstinline{0,0,0,0,b,a,c}.
However,
the sandboxes are \emph{not} in conflict, as the following command show.
\begin{lstlisting}[name=primer]
sbx.inConflictWith(sbx2); // returns false
\end{lstlisting}
While both sandboxes manipulate \lstinline{root}, they manipulate
different fields.
\lstinline{sbx} recalculates the  field \lstinline{value}, whereas
\lstinline{sbx2} replaces the field \lstinline{right}. Neither reads data
that has previously been written by the other sandbox.
However, this situation changes if we call \lstinline{setValue} again,
which also modifies \lstinline{right}.
\begin{lstlisting}[name=primer]
sbx.call(setValue, this, root);*'\label{line:sandboxcall3}'*
var cofts = sbx.conflictsWith(sbx2); // returns a list of conflicts
cofts.foreach(function(i, e) {print(e)});
\end{lstlisting}
It documents a read-after-write conflict:
\begin{lstlisting}
Confict: (1425303937853) get [name=right]@SBX001 - (1425303937855) set [name=right]@SBX002
\end{lstlisting}

%  ___                 _ _      
% / __|___ _ __  _ __ (_) |_ ___
%| (__/ _ \ '  \| '  \| |  _(_-<
% \___\___/_|_|_|_|_|_|_|\__/__/

\subsection{Transaction Processing}

The \emph{commit} operation applies select effects from a sandbox to
its target.
Effects may be committed one at a time by calling \lstinline{commit}
on each effect object or all at once by calling \lstinline{commit} on
the sandbox object.
\begin{lstlisting}[name=primer]
sbx.commit();*'\label{line:commitsbx}'*
root.toString(); // returns 0, 1, 0, 2, 0 
\end{lstlisting}

% ___     _ _ _             _       
%| _ \___| | | |__  __ _ __| |__ ___
%|   / _ \ | | '_ \/ _` / _| / /(_-<
%|_|_\___/_|_|_.__/\__,_\__|_\_\/__/

\noindent
The \emph{rollback} operation undoes an existing manipulation and
returns to its previous configuration before the effect. 
Again, rollbacks can be done on a per-effect basis or for the sandbox
as a whole.
However, a rollback did not remove the shadow object. Thus, after rolling back,
the values are still shadow values in \mbox{\lstinline{sbx}.}
\begin{lstlisting}[name=primer]
sbx.rollback();
root.toString(); // returns 0, 1, 0, 2, 0 
sbx.call(toString, this, root); // returns 0, 0, 0, 0, 0
\end{lstlisting}

% ___                 _   
%| _ \_____ _____ _ _| |_ 
%|   / -_) V / -_) '_|  _|
%|_|_\___|\_/\___|_|  \__|

\noindent
The \emph{revert} operation resets the shadow object of a wrapped value.
The following code snippet reverts the \lstinline{root} object in \lstinline{sbx}.
\begin{lstlisting}[name=primer]
sbx.revertOf(root);
\end{lstlisting}
Now, \lstinline{root}'s shadow object is removed and the origin is
visible again in the sandbox. 
Calling \lstinline{toString} inside of \lstinline{sbx} returns
\lstinline{0,1,0,2,0}. 

%  _____                 _ _               
% / ____|               | | |              
%| (___   __ _ _ __   __| | |__   _____  __
% \___ \ / _` | '_ \ / _` | '_ \ / _ \ \/ /
% ____) | (_| | | | | (_| | |_) | (_) >  < 
%|_____/ \__,_|_| |_|\__,_|_.__/ \___/_/\_\
%                                          
%                                          
% ______                                 _       _   _             
%|  ____|                               | |     | | (_)            
%| |__   _ __   ___ __ _ _ __  ___ _   _| | __ _| |_ _  ___  _ __  
%|  __| | '_ \ / __/ _` | '_ \/ __| | | | |/ _` | __| |/ _ \| '_ \ 
%| |____| | | | (_| (_| | |_) \__ \ |_| | | (_| | |_| | (_) | | | |
%|______|_| |_|\___\__,_| .__/|___/\__,_|_|\__,_|\__|_|\___/|_| |_|
%                       | |                                        
%                       |_|                                        

\section{Sandbox encapsulation}
\label{sec:sandbox}

The implementation of \SBX\ builds on two foundations: \emph{memory safety} and
\emph{reachability}.
In a memory safe programming language, a program cannot access
uninitialized memory or memory outside the range allocated to a datastructure.
An object reference serves as the right to
access the resources managed by the object along with the memory
allocated to it.
In JavaScript, all resources are accessible via property read and write 
operations on objects. Thus, controlling reads and writes is sufficient to control the resources.

\SBX{} ensures isolation of the actual program code by intercepting
each operation that attempts to modify data visible
outside the sandbox. To achieve this behavior, all functions and objects
crossing the sandbox boundary are wrapped in a \emph{membrane} to ensure that the
sandboxed code cannot modify them in any way.
This membrane is implemented using JavaScript proxies \cite{CutsemMiller2010}.

More precisely, our implementation of sandboxing is inspired by \emph{Revocable Membranes}
\cite{CutsemMiller2010,StricklandTobinHochstadtFindlerFlatt2012}
and access control based on object capabilities~\cite{MillerShapiro2003}. 
Identity preserving membranes keep the sandbox apart from the normal program
execution: We encapsulate objects passed through the membrane and
redirect write operations to shadow objects (Section~\ref{sec:sandbox/shadow}),
we encapsulate code (Section~\ref{sec:sandbox/scope}), and we withhold external
bindings from a function (Section~\ref{sec:sandbox/decompile}). No unprotected
value is passed inside the sandbox.

%    _               ___         _      _     ___             _        
% _ | |__ ___ ____ _/ __| __ _ _(_)_ __| |_  | _ \_ _ _____ _(_)___ ___
%| || / _` \ V / _` \__ \/ _| '_| | '_ \  _| |  _/ '_/ _ \ \ / / -_|_-<
% \__/\__,_|\_/\__,_|___/\__|_| |_| .__/\__| |_| |_| \___/_\_\_\___/__/
%                                 |_|                                  

\subsection{Proxies and membranes}
\label{sec:sandbox/proxies}

A \emph{proxy} is an object intended to be used in place of a
\emph{target object}. The proxy's behavior is controlled by a 
\emph{handler object} that typically mediates access to the target
object. Both, target and handler, may be proxy objects themselves. 

The handler object contains trap functions that are called when a
trapped operation is performed on the proxy. Operations like property
read, property write, and function application are forwarded to their
corresponding trap. The trap function may implement the operation arbitrarily,
for example, by forwarding the operation to the target object. The latter is the
default behavior if the trap is not specified.

Figure~\ref{fig:sandbox/proxy} illustrates this situation with a
handler that forwards all operations to the target.

% Set the overall layout of the tree
\tikzstyle{level 0}=[level distance=0em, sibling distance=0em]
\tikzstyle{level 1}=[level distance=8em, sibling distance=2em]
\tikzstyle{level 2}=[level distance=8em, sibling distance=2em]
\tikzstyle{level 3}=[level distance=6em, sibling distance=2em]
\tikzstyle{level 4}=[level distance=4em, sibling distance=2em]
\begin{figure}[t]
  \centering
  \begin{tikzpicture}[-,level/.style={sibling distance=180mm/#1}]
    \node [draw, solid, circle, minimum width=4em, left=1em] (handler) {\textit{Handler}}
    child {%
      node [draw, solid, circle, minimum width=4em, left=1em] (proxy) {\textit{Proxy}}
    }
    child {%
      node [draw, solid, circle, minimum width=4em, right=1em] (target) {\textit{Target}}
    };

    \node[draw=none, anchor=south, below=3em] at (proxy) {%
      \begin{tabular}{l}
        \lstinline!Proxy.x;!\\
        \lstinline!Proxy.x=1;!
      \end{tabular}
    };

    \node[draw=none, anchor=right, right=2em] at (handler) {%
      \begin{tabular}{l}
        \lstinline!Handler.get(Target,'x',Proxy);!\\
        \lstinline!Handler.set(Target,'x',1,Proxy);!
      \end{tabular}
    };

    \node[draw=none, anchor=south, below=3em] (ctarget) at (target) {%
      \begin{tabular}{l}
        \lstinline!Target['x'];!\\
        \lstinline!Target['x']=1;!
      \end{tabular}
    };

    \draw [dashed] (-10em,-4em) -- (14em, -4em)
    node[above, pos=0.92]{\small\textit{Meta-Level}}
    node[below, pos=0.92]{\small\textit{Base-Level}}
    ;

    \node[dashed,rounded corners=7, draw,fit={(target)}] (box) {};
    \path (proxy) edge[dashed] node[]{} (box);

  \end{tikzpicture}
  \caption{Proxy operations. 
    The operation \lstinline{Proxy.x} invokes the trap
    \mbox{\lstinline!Handler.get(Target,'x',Proxy)!} (property get)
    and the property set operation
    \lstinline!Proxy.x=1! invokes
    \mbox{\lstinline!Handler.set(Target,'x',1,Proxy)!}.
  }
  \label{fig:sandbox/proxy}
\end{figure}

% Set the overall layout of the tree
\tikzstyle{level 0}=[level distance=2cm, sibling distance=2cm]
\tikzstyle{level 1}=[level distance=2cm, sibling distance=2cm]
\tikzstyle{level 2}=[level distance=1cm, sibling distance=2cm]
\tikzstyle{level 3}=[level distance=1cm, sibling distance=2cm]
\tikzstyle{level 4}=[level distance=1cm, sibling distance=2cm]
\tikzset{>=latex}
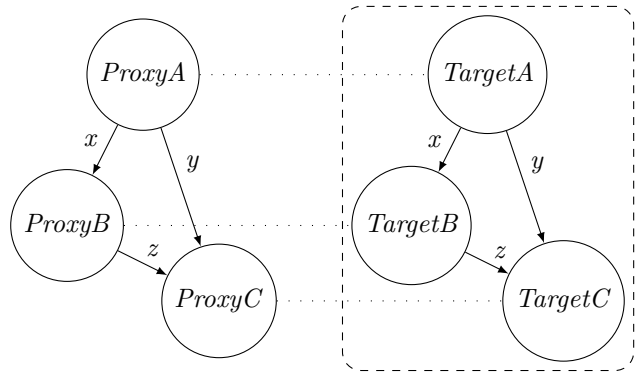
\begin{figure}[t]
  \centering

  \begin{tikzpicture}[-,level/.style={sibling distance=180mm/#1},edge from parent/.style={draw,-latex}]

    \node [left=1.5cm, draw, solid, circle] (root) {\textit{ProxyA}}
    child {%
      node [draw, solid, circle] (rootl) {\textit{ProxyB}}
    }
    child [edge from parent/.style={draw=none}] {%
      node [draw=none, solid, circle] (roodr) {}
      child {%
        node [draw, solid, circle] (rootrr) {\textit{ProxyC}}
      }
    };
    \node [right=1.5cm, draw, draw, circle] (mroot) {\textit{TargetA}}
    child {%
      node [draw, solid, circle] (mrootl) {\textit{TargetB}}
    }
    child [edge from parent/.style={draw=none}] {%
      node [draw=none, solid, circle] (mrootr) {}
      child {%
        node [draw, solid, circle] (mrootrr) {\textit{TargetC}}
      }
    };    
    \path[draw,->] (root) -- (rootrr);
    \path[draw,->] (rootl) -- (rootrr);

    \path[draw,->] (mroot) -- (mrootrr);
    \path[draw,->] (mrootl) -- (mrootrr);

    \path (root) -- (rootl) node[midway, left=1pt, pos=0.3]{\textit{x}};
    \path (root) -- (rootrr) node[midway, right=1pt, pos=0.3]{\textit{y}};
    \path (rootrr) -- (rootl) node[midway, right=7pt, pos=1]{\textit{z}};

    \path (mroot) -- (mrootl) node[midway, left=1pt, pos=0.3]{\textit{x}};
    \path (mroot) -- (mrootrr) node[midway, right=1pt, pos=0.3]{\textit{y}};
    \path (mrootrr) -- (mrootl) node[midway, right=7pt, pos=1]{\textit{z}};

    \node[dashed,rounded corners=7, draw,fit={(mroot) (mrootl) (mrootr) (mrootrr)}] {}; 

    \begin{pgfonlayer}{background}  
      \path[draw,loosely dotted] (root) -- (mroot);
      \path[draw,loosely dotted] (rootl) -- (mrootl);
      \path[draw,loosely dotted] (rootrr) -- (mrootrr);
    \end{pgfonlayer}

  \end{tikzpicture}

  \caption{Property access through an identity preserving membrane
    (dashed line around target objects).
  The property access through the wrapper \lstinline{ProxyA.x}
  returns a wrapper for \lstinline{TargetA.x}. The property access
  \lstinline{ProxyA.y} returns the same wrapper as \lstinline{ProxyB.z}.  
  }
  \label{fig:sandbox/membrane}
\end{figure}

A \emph{membrane} is a regulated communication channel between an object and the
rest of the program. A membrane is implemented by a proxy that guards all operations
on its target. If the result of an operation is another object, then it is
recursively wrapped in a membrane before it is returned. This way, all
objects accessed through an object behind the membrane are also behind the
membrane.
Common use cases of membranes are revoking all references to an object
network at once or enforcing write protection on the objects behind the
membrane~\cite{CutsemMiller2010,Miller2006}.

Figure~\ref{fig:sandbox/membrane} shows a membrane for
\lstinline{TargetA} implemented by wrapper \lstinline{ProxyA}. Each property
access through a wrapper (e.g., \lstinline{ProxyA.x}) returns a
wrapped object. After installing the membrane, no \emph{new} direct
references to target objects behind the membrane become available.

An \emph{identity preserving membrane} guarantees
that no target object has more than one proxy.
% A weak map preserve object identities when creating new wrappers.
Thus, proxy identity outside the membrane
reflects target object identity inside. For example, if 
\lstinline{TargetA.x.z} and \lstinline{TargetA.y} refer to the same object 
(\lstinline{TargetA.x.z===TargetA.y}), then \lstinline{ProxyA.x.z} and
\lstinline{ProxyA.y} refer to the same wrapper object
(\lstinline{ProxyA.x.z===ProxyA.y}).

% ___ _            _               ___  _     _        _      
%/ __| |_  __ _ __| |_____ __ __  / _ \| |__ (_)___ __| |_ ___
%\__ \ ' \/ _` / _` / _ \ V  V / | (_) | '_ \| / -_) _|  _(_-<
%|___/_||_\__,_\__,_\___/\_/\_/   \___/|_.__// \___\__|\__/__/
%                                          |__/               

\subsection{Shadow objects}
\label{sec:sandbox/shadow}

% Set the overall layout of the tree
\tikzstyle{level 0}=[level distance=0em, sibling distance=0em]
\tikzstyle{level 1}=[level distance=8em, sibling distance=2em]
\tikzstyle{level 2}=[level distance=8em, sibling distance=2em]
\tikzstyle{level 3}=[level distance=6em, sibling distance=2em]
\tikzstyle{level 4}=[level distance=4em, sibling distance=2em]

\begin{figure}[t]
  \centering

  \begin{tikzpicture}[-,level/.style={sibling distance=180mm/#1}]
    \node [draw, solid, circle, minimum width=4em, left=1em] (handler) {\textit{Handler}}
    child {%
      node [draw, solid, circle, minimum width=4em, left=-1em] (proxy) {\textit{Proxy}}
    }
    child {%
      node [draw, solid, circle, minimum width=4em, right=1em] (target) {\textit{Target}}
    }
    child {%
      node [draw, solid, circle, minimum width=4em, right=6em] (shadow) {\textit{Shadow}}
    };

    \node[draw=none, anchor=south, below=2em] at (proxy) {%
      \begin{tabular}{l}
        \lstinline!Proxy.x;!\\
        \lstinline!Proxy.y=1;!\\
        \lstinline!Proxy.y;!
      \end{tabular}
    };

    \node[draw=none, anchor=right, right=3em] at (handler) {%
      \begin{tabular}{l}
        \lstinline!Handler.get(Target,'x',Proxy);!\\
        \lstinline!Handler.set(Target,'y',1,Proxy);!\\
        \lstinline!Handler.get(Target,'y',Proxy);!
      \end{tabular}
    };

    \node[draw=none, anchor=south, below=3em] (ctarget) at (target) {%
      \begin{tabular}{l}
        \lstinline!Target['x'];!
      \end{tabular}
    };

    \node[draw=none, anchor=south, below=3em] (cshadow) at (shadow) {%
      \begin{tabular}{l} 
        \lstinline!Shadow['y']=1;!\\
        \lstinline!Shadow['y'];!
      \end{tabular}
    };

    \draw [dashed] (-10em,-4em) -- (14em, -4em)
    node[above, pos=0.92]{\small\textit{Meta-Level}}
    node[below, pos=0.92]{\small\textit{Base-Level}}
    ;

    \node[dashed,rounded corners=7, draw,fit={(target) (shadow)}] (box) {}; 
    \path (proxy) edge[dashed] node[]{} (box);

  \end{tikzpicture}

  \caption{Operations on a sandbox. 
    The property get operation \lstinline{Proxy.x} invokes the trap
    \lstinline{Handler.get(Target,'x',Proxy)}, which forwards the operation to
    the proxy's target. The property set operation \lstinline{Proxy.y=1} invokes
    the trap \lstinline{Handler.set(Target,'y',1,Proxy)}, which forwards the
    operation to a local shadow object. The final property get operation
    \lstinline{Proxy.y} is than also forwarded to the shadow object.
  }
  \label{fig:sandbox/shadow}
\end{figure}
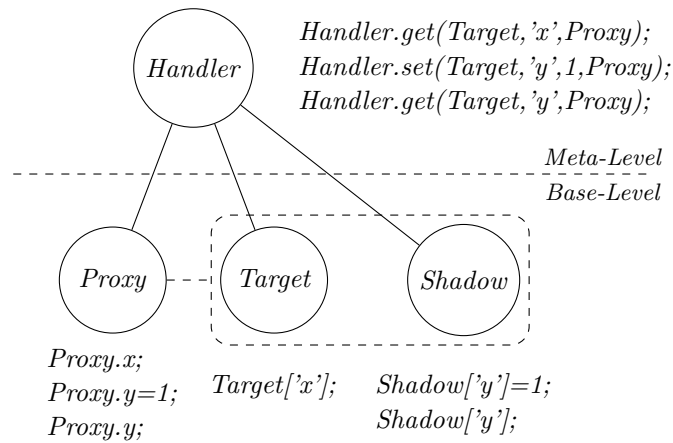

Our sandbox redefines the semantics of proxies to implement
expanders~\cite{WarthStanojevicMillstein2006}, an idea that allows a client side
extension of properties without modifying the proxy's target.

A sandbox handler manages two objects: a target object and a
local \emph{shadow object}. The target object acts as a parent object for its proxy whereas
the shadow object gathers local modifications. Write operations always take
place on the shadow object. A read operation first attempts to obtain
the property from the shadow object. If that fails, the read gets forwarded
to the target object. Figure~\ref{fig:sandbox/shadow} illustrates this
behavior, which is very similar to JavaScript's prototype chain: 
the sandboxed version of an object inherits everything from its outside cousin,
whereas modifications only appear inside the sandbox\footnote{%
  {Getter} and {setter} functions require special
  treatment. Like other functions, they are decompiled and then applied
  to the shadow object. See Section~\ref{sec:sandbox/decompile}.
}.

As sandbox encapsulation extends the functionality of a membrane, each object visible
inside the sandbox is either an object that was created inside or it is a
wrapper for some outside object.

A special proxy wraps sandbox internal values whenever committing a value to the
outside, as shown in the last example. This step mediates uses of a sandbox
internal value in the outside. This is form example required to wrap arguments
values passed to committed sandbox function. The wrapping guarantees that the
sandbox never gets access to unprotected references to the outside.

% ___               _ _               ___                   
%/ __| __ _ _ _  __| | |__  _____ __ / __| __ ___ _ __  ___ 
%\__ \/ _` | ' \/ _` | '_ \/ _ \ \ / \__ \/ _/ _ \ '_ \/ -_)
%|___/\__,_|_||_\__,_|_.__/\___/_\_\ |___/\__\___/ .__/\___|
%                                                |_|        

\subsection{Sandbox scope}
\label{sec:sandbox/scope}

% Set the overall layout of the tree
\tikzstyle{level 0}=[level distance=1em, sibling distance=2em]
\tikzstyle{level 1}=[level distance=1em, sibling distance=2em]
\tikzstyle{level 2}=[level distance=1em, sibling distance=2em]
\tikzstyle{level 3}=[level distance=1em, sibling distance=2em]
\tikzstyle{level 4}=[level distance=1em, sibling distance=2em]

\tikzset{every node/.style={minimum size=0em}}

\tikzset{>=latex}

\tikzset{every fit/.append}
\tikzset{class/.style={%
  rounded corners=1ex 
}, label/.style={%
  align=left,
  node distance=2em,
}}

\begin{figure}[t]
  \centering

  \begin{tikzpicture}
    \node (global) [label, inner sep=0ex, outer sep=0em, align=left, xshift=2ex,
    text width=24em] {%
      \lstinline[keepspaces]!var node = new Node(/* some sub-nodes */);!\\
      \lstinline[keepspaces]!var sbx = new Sandbox(this, /* some parameters */);!\\
    };
    \node (sandbox) [label, inner sep=0ex, outer sep=0ex, align=left, below=of
    global.south, anchor=north, xshift=0em, text width=22em] {%
      \lstinline!with(sbxglobal) {!%
      };

      \node (closure) [label, inner sep=0ex, outer sep=0ex, align=left, below=of
      sandbox.south, anchor=north, xshift=0em, text width=20em] {%
        \lstinline!(function(){!\\
        ~~\lstinline[keepspaces]!``use strict'';!
        };

      \node (setvalue) [label, inner sep=0ex, outer sep=0ex, align=left,
      below=of closure.south, anchor=north, xshift=0em, text width=18em] {%
        \lstinline!function setValue(){!\\
        ~~\lstinline!if(node) {!\\
        ~~~~\lstinline!node.value=heightOf(node);!\\
        ~~~~\lstinline!setValue(node.left);!\\
        ~~~~\lstinline!setValue(node.right);!\\
        ~~\lstinline!}!\\
        \lstinline!}!
      };

      \node (xclosure) [label, inner sep=0ex, outer sep=0ex, align=left,
      below=of setvalue.south, anchor=north, xshift=0em, text width=20em] {%
        \lstinline!})();!%
    };

      \node (xsandbox) [label, inner sep=0ex, outer sep=0ex, align=left,
      below=of xclosure.south, anchor=north, xshift=0em, text width=22em] {%
      \lstinline!}!%
    };
    \node (xglobal) [label, inner sep=0ex, outer sep=0ex, align=left, below=of
    xsandbox.south, anchor=north,  xshift=-0em] {%
    };

    \begin{pgfonlayer}{background}
      \node [fill=gray!50, class, inner sep=1em, outer sep=0em, fit=(global)
      (sandbox) (setvalue) (xsandbox) (xglobal)] {};
      
      \node [draw, rectangle, dashed,
        fill=gray!30, class, inner sep=1em, outer sep=0em, fit=(sandbox)
      (closure) (setvalue) (xclosure) (xsandbox)] {};

      \node [fill=gray!20, class, inner sep=1em, outer sep=0em, fit=(closure)
      (setvalue) (xclosure)] {};
      
      \node [fill=gray!10, class, inner sep=1em, outer sep=0em, fit=(setvalue)]
      {};
    \end{pgfonlayer}

  \end{tikzpicture}

  \caption{Scope chain installed by the sandbox when loading \emph{setValue}.
  The dark box represents the global scope. The dashed line indicates the
  sandbox boundary and the inner box shows the program code nested inside.}
  \label{fig:sandbox/scopechain}
\end{figure}
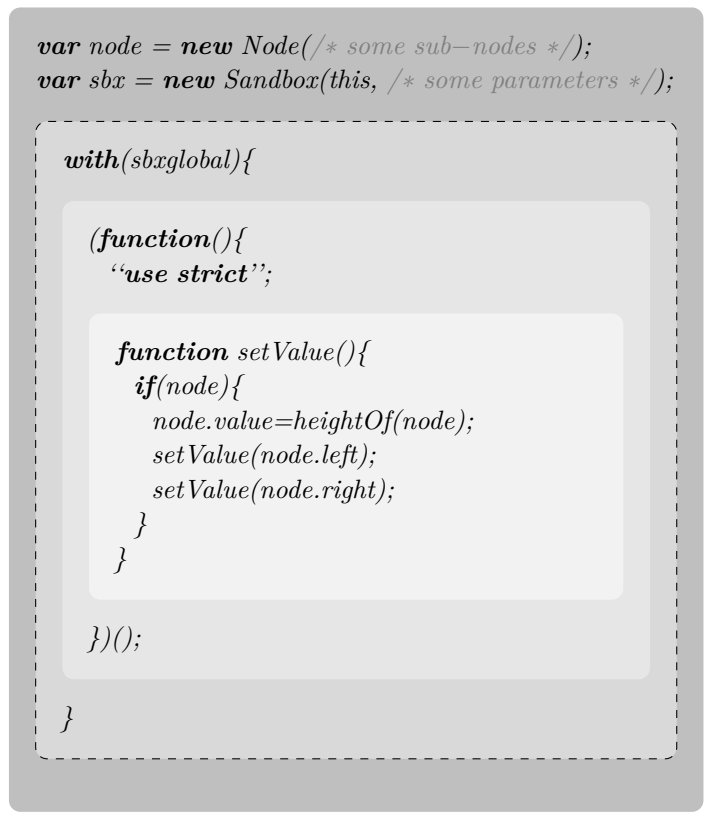

Apart from access restrictions, protecting the global state from
modification through the membrane is fundamental to guarantee noninterference.
To execute program code,
\SBX\ relies on an \lstinline{eval}, which is nested in a statement
\lstinline[keepspaces]!with (sbxglobal) {/* body */}!. The \lstinline{with}
statement places the sandbox global on top of the current
environment's scope chain while 
executing \lstinline{body}. This setup exploits that \lstinline{eval} dynamically rebinds
the free variables of its argument to whatever is in scope at its call
site. In this construction, which is related to dynamic 
binding~\cite{HansonProebsting2001}, any property defined in
\lstinline{sbxglobal} shadows a variable deeper down in the scope chain.

We employ a proxy object in place of \lstinline{sbxglobal} to 
control all non-local variable accesses in the sandboxed code
by trapping the sandbox global object's
\lstinline{hasOwnProperty} method. When JavaScript traverses the scope chain to
resolve a variable access, it calls the method \lstinline{hasOwnProperty} on the
objects of the scope chain starting from the top. Inside the \lstinline{with}
statement, the first object that is checked on this traversal is the proxied
sandbox global. If its \lstinline{hasOwnProperty} method always returns
\lstinline{true}, then the traversal stops here and the JavaScript engine sends
all read and write operations for free variables to the sandbox global. This
way, we obtain full interposition and the handler of the proxied sandbox global
has complete control over the free variables in \lstinline{body}.

Figure~\ref{fig:sandbox/scopechain} visualizes the nested scopes
created during the execution of
\lstinline{setValue} as in the example from Section~\ref{sec:primer}. The sandbox
global \lstinline{sbxglobal} is a wrapper for the actual global object, which is
used to access \lstinline{heightOf} and \lstinline{Math.abs}.
The library code is nested in an empty closure which provides a fresh scope for
local functions and variables. This step is required because JavaScript did not
 have standalone block scopes such as blocks in C or Java.
Variables and named functions
 \footnote{%
   Function created with \lstinline[keepspaces]!function name() \{/* body */\}!.
}  created by the sandboxed code end up in this fresh scope. This extra scope 
guarantees noninterference for dynamically loaded scripts that define
global variables and functions.

The \lstinline{"use strict"}\footnote{%
   Strict mode requires that a use of \lstinline{this} inside a
   function is only valid if either the function was called as a method
   or a receiver object was specified explicitly using
   \lstinline{apply} or \lstinline{call}.
}  declaration in front of the closure puts JavaScript in strict mode,
which ensures that the code cannot obtain unprotected references to
the global object.

% Set the overall layout of the tree
\tikzstyle{level 0}=[level distance=1cm, sibling distance=2cm]
\tikzstyle{level 1}=[level distance=1cm, sibling distance=2cm]
\tikzstyle{level 2}=[level distance=1cm, sibling distance=2cm]
\tikzstyle{level 3}=[level distance=1cm, sibling distance=2cm]
\tikzstyle{level 4}=[level distance=1cm, sibling distance=2cm]

\tikzset{every node/.style={minimum size=0em}}

\tikzset{>=latex}

\tikzset{every fit/.append}
\tikzset{class/.style={%
  rounded corners=1ex 
}, label/.style={%
  align=left,
  node distance=2em,
}}
\tikzset{
  basic box/.style = {
    shape = rectangle,
    align = center,
    fill  = #1!30,
    inner sep=1ex, outer sep=0em,
  rounded corners},
  header node/.style = {
    rounded corners,
    fill=gray!30,
  draw},
  header/.style = {%
    inner ysep = +3ex,
    yshift=2ex,
    append after command = {
      \pgfextra{\let\TikZlastnode\tikzlastnode}
      node [header node, dashed, fill=gray!50, text width=12ex, align = center] (header-\TikZlastnode) at (\TikZlastnode.north) {#1}
    }
  },
}
\pgfdeclarelayer{background1}
\pgfdeclarelayer{background2}
\pgfdeclarelayer{foreground}
\pgfsetlayers{background2,background1,background,main,foreground}   %% some additional layers for demo

\begin{figure}[t]
  \centering

  \begin{tikzpicture}

    \node (scope) [label, inner sep=0ex, outer sep=0em, align=left, xshift=2ex] {%
    };

    \node (global) [label, inner sep=0ex, outer sep=0em, align=center, below=2ex
    of scope.south, anchor=north, xshift=0ex, yshift=2ex, text width=48ex] {%
      Global Scope
    };

    \node (wrapper2) [label, inner sep=0ex, outer sep=0ex, align=left, below=1ex
    of global.south, anchor=north, xshift=0em, text width=12ex, basic box = gray] {%
    };

    \node (wrapper1) [label, inner sep=0ex, outer sep=0ex, align=left, left=3ex
    of wrapper2.west, anchor=east, xshift=0em, text width=12ex, basic box = gray] {%
    };

    \node (wrapper3) [label, inner sep=0ex, outer sep=0ex, align=left, right=3ex
    of wrapper2.east, anchor=west,  xshift=0em, text width=12ex,basic box = gray] {%
    };

    \node (code2) [label, inner sep=0ex, outer sep=0ex, align=left, below=11ex
    of global.south, anchor=north, xshift=0em, text width=12ex, basic box = gray] {%
      JSCode
    };

    \node (code1) [label, inner sep=0ex, outer sep=0ex, align=left, left=3ex of
    code2.west, anchor=east, xshift=0em, text width=12ex, basic box = gray] {%
      JSCode 
    };

    \node (code3) [label, inner sep=0ex, outer sep=0ex, align=left, right=3ex of
    code2.east, anchor=west,  xshift=0em, text width=12ex, basic box = gray] {%
      JSCode
    };

    \begin{pgfonlayer}{background1}
      \node (globalscope) [rectangle, fill=gray!50, class, inner sep=1ex, outer sep=0em, fit=(global)] {};
      \node (sbx1) [draw, rectangle, dashed, fill=gray!10, class, inner sep=1ex, outer sep=0em, fit=(code1), header = Sandbox1] {};
      \node (sbx2) [draw, rectangle, dashed, fill=gray!10, class, inner sep=1ex, outer sep=0em, fit=(code2), header = Sandbox2] {};
      \node (sbx3) [draw, rectangle, dashed, fill=gray!10, class, inner sep=1ex, outer sep=0em, fit=(code3), header = Sandbox3] {};
    \end{pgfonlayer}

    \begin{pgfonlayer}{background2}
      \node (all) [fill=gray!30, class, inner sep=1ex, outer sep=0em, fit=(globalscope) (global) (sbx1) (sbx2) (sbx3)] {};
    \end{pgfonlayer}

    \path[] (code1.north) edge[<->] (header-sbx1.south);
    \path[] (code2.north) edge[<->] (header-sbx2.south);
    \path[] (code3.north) edge[<->] (header-sbx3.south);

    \path[] (wrapper1.north) edge[<->] (header-sbx1.north);
    \path[] (wrapper2.north) edge[<->] (header-sbx2.north);
    \path[] (wrapper3.north) edge[<->] (header-sbx3.north);

  \end{tikzpicture}

  \caption{Nested sandboxes in an application. The outer box represents the
  global application state containing JavaScript's global scope. Each sandbox
  has its own global object and the nested JavaScript code is defined w.r.t.\ to
  the sandbox global.}
  \label{fig:sandbox/scopes}
\end{figure}
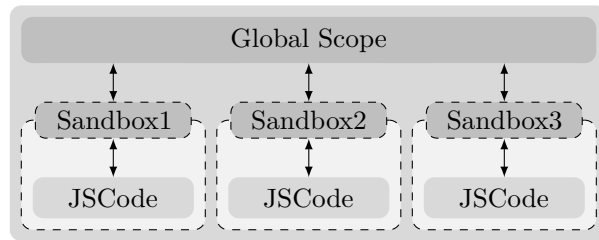

Figure~\ref{fig:sandbox/scopes} shows the situation when instantiating different
sandboxes during program execution. Every sandbox installs its own scope with a  
sandbox global on top of the scope chain. Scripts nested inside are defined with
respect to the sandbox global. The sandbox global mediates the access to
JavaScript's global object. Its default implementation is empty to
guarantee isolation. However, \SBX\ can grant fine-grained access 
by making resources available in the sandbox global.

% ___             _   _          
%| __|  _ _ _  __| |_(_)___ _ _  
%| _| || | ' \/ _|  _| / _ \ ' \ 
%|_| \_,_|_||_\__|\__|_\___/_||_|
%                                
% ___                       _ _      _   _          
%| _ \___ __ ___ _ __  _ __(_) |__ _| |_(_)___ _ _  
%|   / -_) _/ _ \ '  \| '_ \ | / _` |  _| / _ \ ' \ 
%|_|_\___\__\___/_|_|_| .__/_|_\__,_|\__|_\___/_||_|
%                     |_|                           

\subsection{Function recompilation}
\label{sec:sandbox/decompile}

In JavaScript, functions have access to the variables and
functions in the lexical scope in which the function was defined.
The Mozilla documentation\footnote{%
\url{https://developer.mozilla.org/en-US/docs/Web/JavaScript/Closures}
} says: ``It remembers the environment in which it was created.''.
Calls to wrapped functions may still cause side effects through their free
variables (e.g., by modifying a variable or by calling another side-effecting
function). Thus, sandboxing either has to erase external bindings of
functions or it has to verify that a function is free of side effects. The former alternative is
the default in \SBX.

To remove bindings from functions passed through the membrane our
protection mechanism decompiles the function and recompiles it inside 
the sandbox environment. Decompilation relies on the standard implementation of the
\lstinline{toString} method of a JavaScript function that returns a string
containing the source code of the function. Each use of an external
function in a sandbox first decompiles it by calling its
\lstinline{toString} method. To bypass potential tampering, we use a private copy of
\lstinline{Function.prototype.toString} for this call.

Next, we apply \lstinline{eval} to the resulting
string to create a fresh variant of the function.
As explained in Section~\ref{sec:sandbox/scope},
this application of \lstinline{eval} is nested in a \lstinline{with}
statement that supplies the desired environment. Decompilation also
places a \lstinline{"use strict"} statement in front. 

To avoid a frequent decompilation and call of \lstinline{eval} with respect to
the same code, our implementation caches the compiled function where applicable.

Instead of recompiling a function, we may use the string representation of a
function to verify that a function is free of side effects, for example, by checking if the
function's body is SES-compliant\footnote{%
In SES, a function can only cause side effects on values passed as an
argument.}~\cite{SecureEcmaScript}. However, it turns out that recompiling a
function has a lower impact on the execution time than analyzing the function
body.

Functions without a string representation (e.g., native functions like
\lstinline{Object} or \lstinline{Array}) cannot be verified or sanitized before
passing them through the membrane. We can either trust these functions
or rule them out. To this end, \SBX{} may be provided with a white
list of trusted function objects. In any case, functions remain wrapped in a
sandbox proxy to mediate property access.

In addition to normal function and method calls, the access to a property that
is bound to a \lstinline{getter} or \lstinline{setter} function needs to
decompile or verify the \lstinline{getter} or \lstinline{setter} before its
execution.

% ___   ___  __  __   _   _          _      _          
%|   \ / _ \|  \/  | | | | |_ __  __| |__ _| |_ ___ ___
%| |) | (_) | |\/| | | |_| | '_ \/ _` / _` |  _/ -_|_-<
%|___/ \___/|_|  |_|  \___/| .__/\__,_\__,_|\__\___/__/
%                          |_|                         

\subsection{DOM updates}
\label{sec:sandbox/dom}

The \emph{Document Object Model} (DOM) is an API for manipulating HTML
and XML documents that underlie the rendering of a web page. DOM provides a representation of the
document's content and it offers methods for changing its structure, style,
content, etc.
In JavaScript, this API is implemented using special objects,
reachable from the \lstinline{document} object.
% All DOM
% objects come with a range of methods providing access to all aspects of
% the DOM.
Unfortunately, the document tree itself is not an object in the programming
language. Thus, it cannot be wrapped for use inside of a sandbox. The only
possibility is to wrap the interfaces, in particular, the \lstinline{document} object.

We grant access to the DOM by binding the DOM interfaces to the sandbox
global when instantiating a new sandbox. As all interfaces are wrapped in a sandbox
proxy to mediate access, there are a number of
limitations:
\begin{itemize}
  \item By default, DOM nodes are accessed by calling query methods
    like \lstinline{getElementById} on the \lstinline{document} object. 
    Effect logging recognizes these accesses as method calls, rather
    than as operations on the DOM. 
  \item All query functions are special native functions that do not have a
    string representation. Decompilation is not possible so that using
    a query function must be permitted explicitly through the white list.
  \item A query function must be called as a method of an actual DOM object
    implementing the corresponding interface. Thus, DOM objects cannot be
    wrapped like other objects, but they require a special
    wrapping that calls the method on the correct receiver
    object. While read operations can be managed in this way, write
    operations must either be forbidden or they affect the original DOM.
\end{itemize}
Thus, guest code can modify the original DOM unless the DOM
interface is restricted to read-only operations. With unrestricted
operations it would be possible to insert new
\lstinline[language=html]{<script>} elements in the document, which
loads scripts from the internet 
and executes them in the normal application state without further sandboxing.
However, prohibiting write operation means that the majority of
guest codes cannot be executed in the sandbox.

To overcome this limitation, \SBX{} provides guest code access to an
\emph{emulated} DOM instead of the real one. We rely on \emph{dom.js}\footnote{%
  \url{https://github.com/andreasgal/dom.js/}
}, a JavaScript library emulating a full browser DOM, to implement a
DOM interface for scripts running in the sandbox. This emulated DOM
is merged into the global sandbox object when executing scripts.

As this pseudo DOM is constructed inside the sandbox, it can be accessed and
modified at will. No special treatment is required. However, the pseudo DOM is
wrapped in a special membrane mediating all operations and performing
effect logging on all DOM elements. 

As each sandbox owns a direct reference to the sandbox internal DOM it provides
the following features to the user:
% in combination with the followin features
\begin{itemize}

  \item 
    The sandbox provides an interface to the sandbox internal DOM and
    enables the host program to access all aspects of the DOM\@. This
    interface can control the data visible to the guest program. 

  \item A host can load a page template before evaluating guest code.
    This template can be an arbitrary HTML document, like the host's
    page or a blank web page. As most libraries operate on non-blank page
    documents (e.g., by reading or writing to a particular element) this
    template can be used to create an environment.  

  \item Guest code runs without restrictions. For example,
    guest code can introduce new \lstinline[language=html]{<script>}
    elements to load library code from the internet. These libraries are loaded and
    executed inside the sandbox as well. 

  \item All operations on the interface objects are recorded, for example, the access to
    \lstinline{window.location} when loading a document. Effects can examined
    using a suitable API (cf.\ Section~\ref{sec:primer-effects}).

  \item The host program can perform a fine-grained inspection of the
    document tree (e.g., it can search for changes and
    differences). The host recognizes newly created DOM elements
    and it can  transfer content from the sandbox DOM to the DOM of the host program.
\end{itemize}

% ___     _ _    _        
%| _ \___| (_)__(_)___ ___
%|  _/ _ \ | / _| / -_|_-<
%|_| \___/_|_\__|_\___/__/

\subsection{Policies}
\label{sec:sandbox/policies}

A policy is a guideline that prescribes whether an operation is allowed. Most
existing sandbox systems come with a facility to define policies. For
example, a policy may grant access to a certain resource, it may grant
the right to perform an operation or to cause a side effect.

Our system does not provide access control policies in the manner known from
other systems. \SBX\ only provides the mechanism to implement an empty scope and to 
pass selected resources to this scope. When a reference to a certain resource is
made available inside the sandbox, then it should be wrapped in another proxy
membrane that enforces a suitable policy.

For example, one may use this work's transactional membranes to shadow write
operations, \emph{Access Permission Contracts}~\cite{KeilThiemann2013-Proxy} to
restrict the access on objects, or \emph{Revocable
References}~\cite{CutsemMiller2010,CutsemMiller2013} to revoke access to the
outside world.

% _    _       _ _        _   _             
%| |  (_)_ __ (_) |_ __ _| |_(_)___ _ _  ___
%| |__| | '  \| |  _/ _` |  _| / _ \ ' \(_-<
%|____|_|_|_|_|_|\__\__,_|\__|_\___/_||_/__/

\section{Discussion}
\label{sec:limitations}

\paragraph*{Strict Mode}

\SBX\ runs guest code in JavaScript's strict mode to rule out
uncontrolled accesses to the global object. This restriction may lead
to dysfunctional guest code because strict-mode semantics
is subtly different from non-strict mode JavaScript.

However, assuming strict mode is less
restrictive than the restrictions imposed by other techniques that restrict JavaScript's
dynamic features. Alternatively, one could also provide a program transformation
that guards uses of \lstinline{this} that may access the global object.

\paragraph*{Scopes}

\SBX\ places every load in its own scope. Hence,
variables and functions declared in one script are not visible to the
execution of another script in the same sandbox. Indeed, we
deliberately keep scopes apart to avoid interference. To enable communication \SBX\
offers a facility to load mutually dependant scripts into the same
scope. Otherwise, scripts may exchange data by writing to fields in the sandbox
global object.

\paragraph*{Function Decompilation}

If a top level closure is wrapped in a sandbox, then its free valriables have to
be declared to the sandbox or their bindings are removed. Decompilation may
change the meaning of a function, because it rebinds its free variables. Only
``pure functions''\footnote{%
A pure function is a function that only maps its input into an output
without causing any observable side effect.} 
can be decompiled without changing their meaning. 

However, decompiling preserves the semantics of a function if its free variables
are imported in the sandbox. The new closure formed within the sandbox may be
closed over variables defined in that sandbox. This task is rightfully
manual as the availability of global bindings is part of a policy. 

In conclusion, decompilation is unavoidable to guarantee noninterference of a
function defined in another scope as every property read operation may be the
call of a side-effecting getter function. 

\paragraph*{Native Functions}

Decompilation requires a string that contains the source code of that
function, but calling the standard \lstinline{toString} method
from \lstinline{Function.prototype} does not work for all functions.
\begin{itemize}
  \item A native function does not have a string representation. Trust
    in a native function is regulated with a white list of trusted functions.
  \item The \lstinline{Function.prototype.bind()} method creates a new
    function with the same body, but the first couple of arguments
    bound to the arguments of \lstinline{bind()}. JavaScript does not
    provide a string representation for the newly created function.
\end{itemize}

\paragraph*{Object, Array, and Function Initializer}

In JavaScript, some objects can be initialized using a \emph{literal notation}
(initializer notation). Examples are object literals (using \lstinline!{}!),
array objects (using \lstinline{[]}), and function objects (using the named or
unnamed function expression, e.g. \lstinline!function () {}!).
Using the literal notation
circumvents all restrictions and wrappings that we may have imposed on the \lstinline{Object}, \lstinline{Array}, and
\lstinline{Function} constructors.

As we are not able to intercept the construction %, because it is part of the syntax,
using the literal notation enables unprotected read access to the prototype objects
\lstinline{Object.prototype}, \lstinline{Array.prototype}, and
\lstinline{Function.prototype}. The newly created object always inherits from the
corresponding prototype.

However, we will never get access to the prototype object itself and we are
not able to modify the prototype. Writes to the created objects always effect the
object itself and are never forwarded to the prototype object.

Even though all the elements contained in the native prototype objects are
uncritical by default, a global (not sandboxed) script could add sensitive data
or a side effecting function to one of the prototype objects and thus bypass
access to unprotected data.

\paragraph*{Function Constructor}

The function constructor \lstinline!Function!\ creates a new function object
based on the definition given as arguments. 
In contrast to function statements and function expressions, the function
constructor ignores the surrounding scope. The new function is always created in the
global scope and calling it enables access to all global variables. 

To prevent this leakage, the sandbox never grants unwrapped access to
JavaScipt's global \lstinline{Function} constructor even if the 
constructor is white-listed as a safe native function.
A special wrapping intercepts the operations and uses a safe way to construct
a new function with respect to the sandbox.

% _  _          _     _            __                         
%| \| |___ _ _ (_)_ _| |_ ___ _ _ / _|___ _ _ ___ _ _  __ ___ 
%| .` / _ \ ' \| | ' \  _/ -_) '_|  _/ -_) '_/ -_) ' \/ _/ -_)
%|_|\_\___/_||_|_|_||_\__\___|_| |_| \___|_| \___|_||_\__\___|
                                                             
\paragraph*{Noninterference}

The execution of sandboxed code should not interfere with the execution of
application code. That is, the application should run as if no sandboxes were
present. This property is called \emph{noninterference} (NI)
\cite{GoguenMeseguer1982} by the security community. The intuition is that
sandboxed code runs at a lower level of security than application code and that
the low-security sandbox code must not be able to observe the results of the
high-security computation in the global scope.

\SBX\ guarantees integrity and confidentiality.
The default ``empty'' sandbox guarantees to run code in full isolation
from the rest of the application, whereas the sandbox global can provide
protected references to the sandbox.

In summary, the sandboxed code may try to write to an object that is visible to
the application, it may throw an exception, or it may not terminate. 
Our membrane redirects all write operations in sandboxed code to local replicas and it
captures all exceptions. A timeout could be used to transform non-terminating
executions into an exception, alas such a timeout cannot be implemented in
JavaScript.\footnote{%
  The JavaScript \lstinline{timeout} function only schedules a function to run
  when the currently running JavaScript code---presumably some event
  handler---stops. It cannot interrupt a running function.
}

% _____                _   _           _   _____                 _ _       
%|  __ \              | | (_)         | | |  __ \               | | |      
%| |__) | __ __ _  ___| |_ _  ___ __ _| | | |__) |___  ___ _   _| | |_ ___ 
%|  ___/ '__/ _` |/ __| __| |/ __/ _` | | |  _  // _ \/ __| | | | | __/ __|
%| |   | | | (_| | (__| |_| | (_| (_| | | | | \ \  __/\__ \ |_| | | |_\__ \
%|_|   |_|  \__,_|\___|\__|_|\___\__,_|_| |_|  \_\___||___/\__,_|_|\__|___/

\section{Evaluation}
\label{sec:evaluation}

To evaluate our implementation, we applied it to JavaScript benchmark programs
from the Google Octane 2.0 Benchmark
Suite\footnote{\url{https://developers.google.com/octane}}.
These benchmarks
measure a JavaScript engine's performance by running a selection of complex and
demanding programs (benchmark programs run between 5 and 8200 times). 
Google claims that Octane ``measure[s] the performance of
JavaScript code found in large, real-world web applications, running on modern
mobile and desktop browsers.
Each benchmark is complex and demanding .

As expected, the run time increases when executing a benchmark in a sandbox.
While some programs like \emph{EarleyBoyer}, \emph{NavierStrokes},
\emph{pdf.js}, \emph{Mandreel}, and \emph{Box2DWeb} are heavily affected, others
are only slightly affected: \emph{Richards}, \emph{Crypto}, \emph{RegExp}, and
\emph{Code loading}, for instance. The observed run time impact entirely depends
on the number of values that cross the membrane.

From the running times we find that the sandbox itself causes an average
slowdown of 8.01 (over all benchmarks). This is more than acceptable compared to
other language-embedded systems.
The numbers also show that sandboxing with fine-grained effect logging enabled
causes an average slowdown of 32.60, an additional factor of 4.07 on top of pure
sandboxing.

Because the execution of program code inside of a sandbox is nothing else than a
normal program execution inside of a \lstinline{with} statement and with
one additional call to \lstinline{eval} (when instantiating the execution) the
run-time impact is influenced by
\begin{inparaenum}[(i)]
  \item the number of wrap operations of values that cross the membrane,
  \item the number of decompile operations on functions, and
  \item the number of effects on wrapped objects.
\end{inparaenum}
Readouts from internal counters indicate that the heavily affected benchmarks
(\emph{RayTrace}, \emph{pdf.js}, \emph{Mandreel}, and \emph{Box2DWeb}) perform a
very large number of effects. The \emph{RayTrace} benchmark, for example,
performs 51~million effects.

Overall, an average slowdown of 8.01 is more than acceptable
compared to other language-embedded systems.
As Octane is intended to measure the engine's performance (benchmark programs
run between 5 and 8200 times) we claim that it is the heaviest kind of
benchmark. Every real-world library (e.g.\ \lstinline{jQuery}) is less demanding
and runs without an measurable runtime impact.

Appendix~\ref{sec:appendix/evaluation} also contains the score values
obtained from running the benchmark suite and lists the readouts of some
internal counters.

%  _____                 _           _             
% / ____|               | |         (_)            
%| |     ___  _ __   ___| |_   _ ___ _  ___  _ __  
%| |    / _ \| '_ \ / __| | | | / __| |/ _ \| '_ \ 
%| |___| (_) | | | | (__| | |_| \__ \ | (_) | | | |
% \_____\___/|_| |_|\___|_|\__,_|___/_|\___/|_| |_|

\section{Conclusion}
\label{sec:conclusion}

\SBX\ runs JavaScript code in a configurable degree of isolation with
fine-grained access control rather than disallowing all access to the
application state. 
It provides full browser compatibility (i.e. all browsers
work without modifications as long as the proxy API is supported) and it has a
better performance than other language-embedded systems.

Additionally, \SBX\ comes with the following features:
\begin{enumerate}
  \item \emph{Language-embedded sandbox.}
    \SBX{} is a JavaScript library and all aspects are
    accessible through a sandbox API\@. The library can be deployed as a language
    extension and requires no changes in the JavaScript run-time system.
  \item \emph{Full interposition.}
    \SBX{} is implemented using Java\-Script proxies~\cite{CutsemMiller2010}.
    The proxy-based implementation guarantees full interposition for the full
    JavaScript language including all dynamic
    features (e.g., \lstinline{with}, \lstinline{eval}).
    \SBX{} works for all code
    regardless of its origin, including dynamically loaded code and code
    injected via \lstinline{eval}.
    No source code transformation or avoidance of
    JavaScript's dynamic features is required.
  \item \emph{Transaction-based sandboxing.}
    A \SBX\ sandbox provides a transactional scope that logs all effects. 
    Wrapper proxies make external
    objects accessible inside of the sandbox and enable sandbox internal
    modifications of the object. Hence, sandboxed code runs as
    usual without noticing the sandbox.
    Effects reveal conflicts,
    differences, and changes with respect to another sandbox or the global
    state. After inspection of the log, effects can be committed to the
    application state or rolled back.
\end{enumerate}

\subparagraph*{Acknowledgments}
This work benefited from discussions with participants of the Dagstuhl Seminar
``Scripting Languages and Frameworks: Analysis and Verification'' in 2014. In
particular, Tom Van Cutsem provided helpful advice on the internals of
JavaScript proxies.

%%
%% Bibliography
%%

%\bibliography{abbrevs,papers,theses,misc,collections,books}
%\bibliography{main}

%%
%% Appendix
%%

\newpage
\appendix
% _____      _                     
%|  __ \    (_)                    
%| |__) | __ _ _ __ ___   ___ _ __ 
%|  ___/ '__| | '_ ` _ \ / _ \ '__|
%| |   | |  | | | | | | |  __/ |   
%|_|   |_|  |_|_| |_| |_|\___|_|   
                                  
\section{Motivation}
\label{sec:appendix/motivation}

JavaScript is the most important client side language for
web pages.
JavaScript developers rely heavily on
third-party libraries for calenders, maps, social networking, feature
extensions, and so on. Thus, the client-side code of a web page is usually composed of
dynamically loaded fragments from different origins.

However, the JavaScript language has no provision for namespaces or
encapsulation management: there is a global scope for variables 
and functions, and every loaded script has the same authority.
On the one hand, JavaScript developers benefit from JavaScript's flexibility as it enables to extend the
application state easily. On the other hand, once included, a script has the ability to
access and manipulate every value reachable from the global
object. That makes it difficult to enforce any security policy in JavaScript.

As a consequence, program understanding and maintenance becomes very
difficult because side effects may cause unexpected behavior. There
is also a number of security concerns as the library code
may access sensitive data, for example, it may read user input from the
browser's DOM.

Browsers normally provide build-in isolation mechanisms. However, as isolation
is not always possible for all scrips, the key challenges of a JavaScript
developer is to manage untrusted third-party code, to control the use of data by
included scrips, and to reason about effects of included code.

\subsection{JavaScript issues}
\label{sec:appendix/motivation/issues}

\begin{figure}[t]
  \centering
\begin{lstlisting}[name=appendix/motivation, language=HTML5]
<!DOCTYPE html>
  <html lang="en">
  <head>
    <!-- third-party libraries -->
    <script src="date.js"></script>
    <script src="jquery.js"></script>
    <script src="jquery.formatDateTime.js"></script> 
  </head>
  <body>
    <!-- Body of the page -->
    <h1 id="headline">Headline</h1>
    <script type="text/javascript">
      window.*'\$'*("*'\#'*headline").text("Changed Headline");
    </script>
  </body>
</html>
\end{lstlisting} 
\caption{Motivating Example. The listing shows a snippet of an 
  \emph{index.html} file. The \lstinline[language=HTML5]{<script>} tags load
  third-party libraries to the application state before executing the body.
  Within the \lstinline[language=HTML5]{<body>} tag it uses \emph{jQuery} to
  modify the DOM.}
  \label{fig:appendix/motivation/normal}
\end{figure}

As an example, we consider a web
application that relies on third-party scripts from various
sources.
Figure~\ref{fig:appendix/motivation/normal} shows an extract of such a page. It first
includes \emph{Datejs}\footnote{%
  \url{https://github.com/datejs/Datejs}
}, a library extending JavaScript's native \lstinline{Date} object with
additional methods for parsing, formatting, and processing of dates. Next, it
loads \emph{jQuery}\footnote{%
  \url{https://jquery.com/}
} and a \emph{jQuery} plugin \emph{jquery.formatDateTime.js}\footnote{%
  \url{https://github.com/agschwender/jquery.formatDateTime}
} that also simplifies formatting of JavaScript date objects.

At this point, we want to ensure that loading the third-party code
(\emph{Datejs} and \emph{jQuery}) does not influence the application state in an
unintended way. Encapsulating the library code in a sandbox enables us to
scrutinize modifications that the foreign code may attempt and only commit
acceptable modifications.

% ___         _      _   _           
%|_ _|___ ___| |__ _| |_(_)_ _  __ _ 
% | |(_-</ _ \ / _` |  _| | ' \/ _` |
%|___/__/\___/_\__,_|\__|_|_||_\__, |
%                              |___/ 
% _   _    _        _                    _        
%| |_| |_ (_)_ _ __| |___ _ __  __ _ _ _| |_ _  _ 
%|  _| ' \| | '_/ _` |___| '_ \/ _` | '_|  _| || |
% \__|_||_|_|_| \__,_|   | .__/\__,_|_|  \__|\_, |
%                        |_|                 |__/ 
%    _               ___         _      _   
% _ | |__ ___ ____ _/ __| __ _ _(_)_ __| |_ 
%| || / _` \ V / _` \__ \/ _| '_| | '_ \  _|
% \__/\__,_|\_/\__,_|___/\__|_| |_| .__/\__|
%                                 |_|       

\subsection{Isolating third-party JavaScript}
\label{sec:appendix/motivation/isolation}

\begin{figure}[t]
  \centering
\begin{lstlisting}[name=appendix/motivation, language=HTML5]
<!DOCTYPE html>
  <html lang="en">
  <head>
    <!-- DecentJS code-->
    <script src="decent.js"></script>
    <!-- Runs Datejs in a fresh sandbox. -->
    <script type="text/javascript">
      var sbx = new Sandbox(this, Sandbox.DEFAULT);*'\label{line:appendix/motivation/sbx}'*
      sbx.request("datejs.js");*'\label{line:appendix/motivation/sbx/datejs}'*
      sbx.applyRule(*'\label{line:appendix/motivation/sbx/rule}'*
        new Rule.CommitOn(Date, function(sbx, effect) {
          return (effect instanceof Effect.Set) &&
            !(effect.name in Date);
      }));
    </script>
    <!-- ... -->
    </head>
  <body>
    <!-- ... -->
  </body>
</html>
\end{lstlisting}
\caption{Execution of library code in a sandbox. The first
  \lstinline[language=HTML5]{<script>} tag loads the sandbox implementation. The
  body of the second \lstinline[language=HTML5]{<script>} tag instantiates a new
  sandbox and loads and executes \emph{Datejs} inside the sandbox. Later it
  commits   intended effects to the native \lstinline{Date} object.}
  \label{fig:appendix/motivation/sandbox}
\end{figure}

Transactional sandboxing is inspired by the idea of transaction processing in
database systems~\cite{WeikumVossen2001:_trans} and software transactional
memory~\cite{ShavitTouitou1995}. Each sandbox implements a transactional scope
the content of which can be examined, committed, or rolled back.

\begin{enumerate}
  \item \emph{Isolation of code.} A \SBX{} sandbox can run JavaScript code in
    isolation to the application state. Proxies make external values visible
    inside of the sandbox and handle sandbox internal write operations. An
    internal DOM simulates the browser DOM as needed. This setup guarantees that the
    isolated code runs without noticing the sandbox.
  \item \emph{Providing transactional features.} A \SBX{} sandbox
    provides a transactional scope in which effects are logged for
    inspection. Policy rules can be specified so that only 
    effects that adhere to the rules are committed to the application
    state and others are rolled back.
\end{enumerate}
Appendix~\ref{sec:primer} gives a detailed introduction to \SBX's API and
provides a series of examples explaining its facilities.

Figure~\ref{fig:appendix/motivation/sandbox} shows how to modify the \emph{index.html} from
Figure~\ref{fig:appendix/motivation/normal} to load the third-party code into a sandbox. We
first focus on \emph{Datejs} and consider \emph{jQuery} later in
Section~\ref{sec:appendix/motivation/dom}. The \lstinline{<!-- ... -->} comment is a
placeholder for unmodified code not considered in this
example\footnote{Appendix~\ref{sec:appendix/appendix/motivation/full} shows the full HTML
code.}.

Initially, we create a fresh sandbox (line~\ref{line:appendix/motivation/sbx}). The first
parameter is the sandbox-internal global object for scripts running in the
sandbox whereas the second parameter is a configuration object\footnote{%
\lstinline{Sandbox.DEFAULT} is a predefined configuration object for the
standard use of the sandbox.
It consists of simple key-value pairs, e.g.\
\lstinline{verbose:false}.
}.

The sandbox global object acts as a mediator between the sandbox
contents and the
external world (cf.\ Section~\ref{sec:sandbox/scope}). It is placed on top of
the scope chain for code executing inside the sandbox and it can be used to make
outside values available inside the sandbox. It is wrapped in a proxy membrane
to mediate all accesses to the host program.

Next, we instruct the sandbox to load and execute the \emph{Datejs} library
(line~\ref{line:appendix/motivation/sbx/datejs}) inside the sandbox. Afterwards,
the sandbox-internal proxy for JavaScript's native \lstinline{Date}
object is 
modified in several ways. Among others, the library adds new methods 
to the \lstinline{Date} object and extends \lstinline{Date.prototype} with
additional properties. Write operations on a proxy wrapper produce a \emph{shadow value}
(cf.\ Section~\ref{sec:sandbox/shadow}) that represents the sandbox-internal
modification of an object. Initially, this modification is only visible to reads
inside the sandbox. Reads are forwarded to the target unless there are local
modifications, in which case the shadow value is returned. The return values are
wrapped in proxies, again.

%  ___                 _ _   _   _             _     _               _        _ 
% / __|___ _ __  _ __ (_) |_| |_(_)_ _  __ _  (_)_ _| |_ ___ _ _  __| |___ __| |
%| (__/ _ \ '  \| '  \| |  _|  _| | ' \/ _` | | | ' \  _/ -_) ' \/ _` / -_) _` |
% \___\___/_|_|_|_|_|_|_|\__|\__|_|_||_\__, | |_|_||_\__\___|_||_\__,_\___\__,_|
%                                      |___/                                    
% __  __         _ _  __ _         _   _             
%|  \/  |___  __| (_)/ _(_)__ __ _| |_(_)___ _ _  ___
%| |\/| / _ \/ _` | |  _| / _/ _` |  _| / _ \ ' \(_-<
%|_|  |_\___/\__,_|_|_| |_\__\__,_|\__|_\___/_||_/__/
                                                    
\subsection{Committing intended modifications}
\label{sec:appendix/motivation/commit}

During execution, each sandbox records the effects on all objects that cross the
sandbox membrane\footnote{%
  The lists do not contain effects on values that were created inside of the
  sandbox.
}. The sandbox API offers access to the resulting lists for inspection and
provides query methods to select the effects of a particular object. After
loading \emph{Datejs}, the effect log reports 16 reads and 142 writes on three
different objects\footnote{%
  Appendix~\ref{sec:appendix/appendix/motivation/effects} shows a readout of the effect lists.
}. However, as the manual inspection of effects is impractical
and requires a lot of effort, \SBX\ allows us to register \emph{rules} with a
sandbox and apply them automatically. A
rule combines a sandbox operation with a predicate specifying the state under
which the operation is allowed to be performed.

For example, as we consider an extension to the 
\lstinline{Date} object as intended, non-critical  modification, we
install a rule that automatically commits new properties to the
\lstinline{Date} object in 
Line~\ref{line:appendix/motivation/sbx/rule}.
In general, a rule \lstinline!CommitOn! takes a target object
(\lstinline{Date}) and a predicate. The predicate function gets invoked with the \emph{sandbox
object} (\lstinline{sbx}) and an \emph{effect object} describing an
effect on the target object. In
our example, the predicate checks if the effect is a property write operation
extending JavaScript's native \lstinline{Date} object and that the 
property name is not already present.

If we construct a function inside of a sandbox and this function
is written and committed to an outside object, then the free variables
of the function contain objects inside the sandbox and arguments of a
call to this function are also wrapped. That is, calling
this function on the outside only causes effects inside the sandbox.
Furthermore, committing an object in this way wraps the object in a
proxy before writing it to its (outside) target. Both measures are required
to guarantee that the sandbox never gets access to  unwrapped
references from the outside world.

At this point we have to mention that the data structure of the committed
functions is constructed inside of \lstinline{sbx}. All bound references
of those functions still point to objects inside of the sandbox and thus using
them only causes effect inside of the sandbox. Furthermore, committing an
object wraps the object in a proxy before writing it to its target. This
intercepts the use of the committed object, e.g.\ to wrap the
arguments of a committed function before invoking the function. This is

% Set the overall layout of the tree
\tikzstyle{level 0}=[level distance=0ex, sibling distance=10ex]
\tikzstyle{level 1}=[level distance=10ex, sibling distance=10ex]
\tikzstyle{level 2}=[level distance=10ex, sibling distance=12ex]
\tikzstyle{level 3}=[level distance=8ex, sibling distance=10ex]
\tikzstyle{level 4}=[level distance=8ex, sibling distance=10ex]

\tikzset{every node/.style={minimum size=2em}}

\tikzset{%
  invisible/.style={opacity=0},
}

\begin{figure}[t]
  \centering
  \begin{tikzpicture}[-,edge from parent/.style={draw,-latex}]
    \node [left=10ex, draw, solid] (date) {\textit{Date}}
    child {%
      node [solid, right=0em] (indate) {\textit{$\cdots$}}
    }
    child {%
      node [draw, solid] (now) {\textit{now()}}
    }
    child {%
      node [draw, solid] (prototype) {\textit{prototype}}
      child {%
        node [draw, solid] (tostring) {\textit{toString()}}
      }
      child {%
        node [solid, left=0em] (rightprototype) {\textit{$\cdots$}}
      }
    };

    \node [right=10ex, draw, draw, dashed] (sbxdate) {\textit{Date}}
    child[dashed] {%
      node [draw, solid, invisible] (sbxnow) {\textit{now()}}
      edge from parent[draw=none]
    }
    child[dashed] {%
      node [draw, dashed] (sbxprototype) {\textit{prototype}}
      child[dashed] {%
        node [draw, solid, invisible] (sbxtostring) {\textit{toString()}}
        edge from parent[draw=none]
      }
      child[dashed] {%
        node [draw, solid, invisible] (sbxisweekday) {\textit{isWeekday()}}
        edge from parent[draw=none]
      }
    };

    \node[dashed,rounded corners=7,draw,fit={(sbxdate) (sbxtostring) (sbxisweekday) (sbxnow)}, inner sep=1ex] {}; 

    \begin{pgfonlayer}{background}  
      %\path[draw,loosely dotted] (sbxdate) -- (date);
      %\path[draw,loosely dotted] (sbxprototype) -- (sbxnow);
      %\path[draw,loosely dotted] (sbxnow) -- (prototype);

      \path[draw,dotted] (sbxdate) edge [bend right=10,->] (date);
      %\path[draw,dotted] (now) edge [bend left=40,->] (sbxnow);
      \path[draw,dotted] (sbxprototype) edge [bend right=10,->] (prototype);
    \end{pgfonlayer}

  \end{tikzpicture}

  \vspace{0.5\baselineskip}
  \hrule
  \vspace{0.5\baselineskip}

  \begin{tikzpicture}[-,edge from parent/.style={draw,-latex}]
    \node [left=10ex, draw, solid] (date) {\textit{Date}}
    child {%
      node [solid, right=0em] (indate) {\textit{$\cdots$}}
    }
    child {%
      node [draw, solid] (now) {\textit{now()}}
    }
    child {%
      node [draw, solid] (prototype) {\textit{prototype}}
      child {%
        node [draw, solid] (tostring) {\textit{toString()}}
      }
      child {%
        node [solid, left=0em] (rightprototype) {\textit{$\cdots$}}
      }
    };

    \node [right=10ex, draw, draw, dashed] (sbxdate) {\textit{Date}}
     child[dashed] {%
      node [draw, solid] (sbxnow) {\textit{now()}}
      edge from parent[solid]
    }
    child[dashed] {%
      node [draw, dashed] (sbxprototype) {\textit{prototype}}
      child[dashed] {%
        node [draw, solid] (sbxtostring) {\textit{toString()}}
        edge from parent[solid]
      }
      child[dashed] {%
        node [draw, solid] (sbxisweekday) {\textit{isWeekday()}}
        edge from parent[solid]
      }
    };

    \node[dashed,rounded corners=7,draw,fit={(sbxdate) (sbxtostring) (sbxisweekday) (sbxnow)}, inner sep=1ex] {}; 

    \begin{pgfonlayer}{background}  
      %\path[draw,loosely dotted] (sbxdate) -- (date);
      %\path[draw,loosely dotted] (sbxprototype) -- (sbxnow);
      %\path[draw,loosely dotted] (sbxnow) -- (prototype);

      \path[draw,dotted] (sbxdate) edge [bend right=10,->] (date);
      %\path[draw,dotted] (now) edge [bend left=40,->] (sbxnow);
      \path[draw,dotted] (sbxprototype) edge [bend right=40,->] (prototype);
    \end{pgfonlayer}

  \end{tikzpicture}

  \vspace{0.5\baselineskip}
  \hrule
  \vspace{0.5\baselineskip}

  \begin{tikzpicture}[-,edge from parent/.style={draw,-latex}]
    \node [left=10ex, draw, solid] (date) {\textit{Date}}
    child {%
      node [solid, right=0em] (indate) {\textit{$\cdots$}}
    }
    child {%
      node [draw, dashed] (now) {\textit{now()}}
      edge from parent[dashed]
    } 
    child {%
      node [draw, solid] (prototype) {\textit{prototype}}
      child {%
        node [draw, solid] (tostring) {\textit{toString()}}
      }
      child {%
        node [solid, left=0em] (rightprototype) {\textit{$\cdots$}}
      }
    };

    \node [right=10ex, draw, draw, dashed] (sbxdate) {\textit{Date}}
    child[dashed] {%
      node [draw, solid] (sbxnow) {\textit{now()}}
      edge from parent[solid]
    }
    child[dashed] {%
      node [draw, dashed] (sbxprototype) {\textit{prototype}}
      child[dashed] {%
        node [draw, solid] (sbxtostring) {\textit{toString()}}
        edge from parent[solid]
      }
      child[dashed] {%
        node [draw, solid] (sbxisweekday) {\textit{isWeekday()}}
        edge from parent[solid]
      }
    };

    \node[dashed,rounded corners=7,draw,fit={(sbxdate) (sbxtostring) (sbxisweekday) (sbxnow)}, inner sep=1ex] {}; 

    \begin{pgfonlayer}{background}
      %\path[draw,loosely dotted] (sbxdate) -- (date);
      %\path[draw,loosely dotted] (sbxprototype) -- (sbxnow);
      %\path[draw,loosely dotted] (sbxnow) -- (prototype);
      %\path[draw,loosely dotted] (prototype) -- (now);

      \path[draw,dotted] (sbxdate) edge [bend right=10,->] (date);
      \path[draw,dotted] (now) edge [bend left=40,->] (sbxnow);
      \path[draw,dotted] (sbxprototype) edge [bend right=40,->] (prototype);
      
    \end{pgfonlayer}

  \end{tikzpicture}
  \caption{%
    Shadow objects in the sandbox when loading \emph{Datejs}
    (cf.\ Section~\ref{sec:primer/isolation}). The structure of JavaScrip's
    native \lstinline{Date} object is shown in solid lines on the left. 
    The shadow values are enclosed by a dashed line. 
    Solid lines are direct references to non-proxy objects, whereas dashed lines
    are indirect references and proxy objects. Dotted lines connect to
    the target object.
    The first box shows the sandbox
    after reading \lstinline{Date.prototype} whereas the second box shows
    the sandbox after modifying the structure of \lstinline{Date}.
    The third box shows the situation after committing the modifications on
    \lstinline{Date}.
  }
  \label{fig:appendix/motivation/example/tree}
\end{figure}
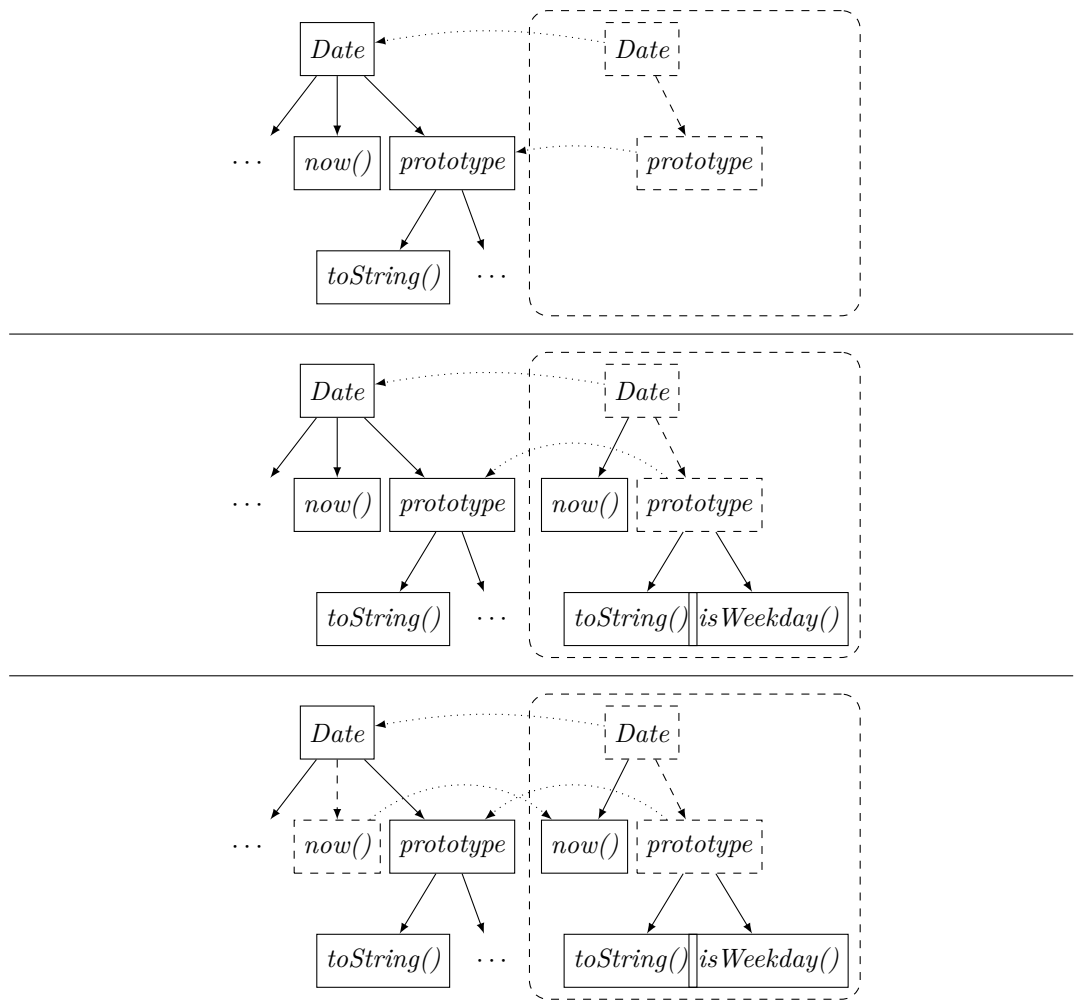

As an illustration, Figure~\ref{fig:appendix/motivation/example/tree} shows an extract of the
membrane arising from JavaScript's native \lstinline{Date} object in
Appendix~\ref{sec:appendix/motivation/}.
Executing \emph{Datejs} in \lstinline{sbx} (shown on the left in the
first box) creates a proxy
for each element accessed on \lstinline{Date}: \lstinline{Date} and
\lstinline{Date.prototype}. Only \lstinline{Date} and \lstinline{Date.prototype}
are wrapped because proxies are created on demand. As proxies forward
each read to the target the structure visible inside of the sandbox is
identical to the structure visible outside.

Extending the native \lstinline{Date} object in \lstinline{sbx} yields the state
shown in the second box. All modifications are only visible inside of the
sandbox. The new elements are not wrapped because they only exist inside of the
sandbox.

However, a special proxy wraps sandbox internal values whenever committing a
value to the outside, as shown in the last box. This step mediates
further uses of the sandbox internal value, for example, wrapping the \lstinline{this}
value and all arguments when calling a function defined in the sandbox.
The wrapping guarantees that the sandbox never gets access to unprotected references to the
outside.

% ___ _            _            _             ___   ___  __  __ 
%/ __| |_  __ _ __| |_____ __ _(_)_ _  __ _  |   \ / _ \|  \/  |
%\__ \ ' \/ _` / _` / _ \ V  V / | ' \/ _` | | |) | (_) | |\/| |
%|___/_||_\__,_\__,_\___/\_/\_/|_|_||_\__, | |___/ \___/|_|  |_|
%                                     |___/                     
%  ___                     _   _             
% / _ \ _ __  ___ _ _ __ _| |_(_)___ _ _  ___
%| (_) | '_ \/ -_) '_/ _` |  _| / _ \ ' \(_-<
% \___/| .__/\___|_| \__,_|\__|_\___/_||_/__/
%      |_|                                   

\subsection{Shadowing DOM operations}
\label{sec:appendix/motivation/dom}

\begin{figure}[t]
  \centering
\begin{lstlisting}[name=appendix/motivation, language=HTML5]
<!DOCTYPE html>
  <html lang="en">
  <head>
    <!-- DecentJS code-->
    <script src="decent.js"></script>
    <!-- ... -->
    <!-- Runs jQuery in a fresh sandbox. -->
    <script type="text/javascript">
      var sbx2 = new Sandbox(this, Sandbox.WEB);*'\label{line:appendix/motivation/sbx2}'*
      sbx2.initialize("template.html");*'\label{line:appendix/motivation/sbx2/template}'*
      sbx2.applyRule(new Rule.Commit(this, "jQuery"));*'\label{line:appendix/motivation/sbx2/rule1}'*
      sbx2.applyRule(new Rule.Commit(this, "*'\$'*"));*'\label{line:appendix/motivation/sbx2/rule2}'*
    </script>
    </head>
  <body>
    <!-- Body of the page -->
    <h1 id="headline">Headline</h1>
    <script type="text/javascript">
      window.*'\$'*("*'\#'*headline").text("Changed Headline");*'\label{line:appendix/motivation/sbx2/jquery}'*
      document.getElementById("headline").innerHTML=sbx2.dom.document.getElementById("headline").innerHTML;*'\label{line:appendix/motivation/sbx2/dom}'*
    </script>
  </body>
</html>
\end{lstlisting}
\caption{Execution of a web library in a sandbox. The first
  \lstinline[language=HTML5]{<script>} tag loads the sandbox implementation. The
  body of the second \lstinline[language=HTML5]{<script>} tag instantiates a new
  sandbox and initializes the sandbox with a predefined HTML template (see
  Figure~\ref{fig:appendix/motivation/template}). Later it commits intended effects to the
application state and copies data from the sandbox internal DOM.}
  \label{fig:appendix/motivation/withdom}
\end{figure}

\begin{figure}[t]
  \centering
\begin{lstlisting}[name=appendix/motivation, language=HTML5]
<!DOCTYPE html>
  <html lang="en">
  <head>
    <script src="jquery.js"></script>
    <script src="jquery.formatDateTime.js"></script> 
  </head>
  <body>
    <!-- Body of the page -->
    <h1 id="headline">Headline</h1>
  </body>
</html>
\end{lstlisting} 
  \caption{File \emph{template.html} contains the
  \lstinline[language=HTML5]{<script>} tags for loading the \emph{jQuery} code
  from \emph{index.html} in Figure~\ref{fig:appendix/motivation/normal}.}
  \label{fig:appendix/motivation/template}
\end{figure}

The example in Figure~\ref{fig:appendix/motivation/sandbox} omits the inclusion of
\emph{jQuery} for simplification purposes.
However, our initial objective is to sandbox all third-party code to \begin{inparaenum}[i]
\item reason about the modifications done by loading the third-party code
\item prevent the application state from unintended modifications
\end{inparaenum}. 

Isolating a library like \emph{jQuery} is more challenging as it
needs access to the browser's DOM. Calls to the native DOM interface
expose a mixture of public and confidential information, so the access
can neither be fully trusted nor completely
forbidden. To address this issue, \SBX\ provides an \emph{internal DOM}\footnote{%
  See Section~\ref{sec:sandbox/dom} for a more detailed discussion.
} that serves as a shadow for the actual DOM when running a web library in the
sandbox. 

Figure~\ref{fig:appendix/motivation/withdom} demonstrates loading the \lstinline{jQuery}
library in a web sandbox. As we extend the first example, we create a new empty sandbox
(line~\ref{line:appendix/motivation/sbx2}) and initialize the sandbox internal DOM by
loading an HTML template (line~\ref{line:appendix/motivation/sbx2/template}). 
Using the \lstinline{Sandbox.WEB}
configuration activates the shadow DOM by instructing the sandbox to create a DOM interface and to merge
this interface with the sandbox-internal global object. The shadow DOM
initially contains an empty document.
It can be instantiated with the actual HTML body or with an HTML
template, as shown in 
Line~\ref{line:appendix/motivation/sbx2/template}.

Figure~\ref{fig:appendix/motivation/template} shows the template, which is an
extract of the original \emph{index.html} containing only the
\lstinline{<script>} tags for the \emph{jQuery} library and selected parts of
the HTML body.
Loading the template also loads and executes the third-party code inside the
sandbox. Afterwards, the internal effect log reports two write operations to the
fields \emph{\$} and \emph{jQuery} of the global \lstinline{window} object, and
one write operation to the \lstinline{HTMLBodyElement} interface, a child of
\lstinline{Node}, both of which are part of the DOM interface.

To automatically commit intended modifications to the
global \lstinline{window} object, we install a suitable rule in Lines~\ref{line:appendix/motivation/sbx2/rule1}
and~\ref{line:appendix/motivation/sbx2/rule2}. As \emph{jQuery} has been
instantiated w.r.t.\ the sandbox internal DOM\@, using it modifies the
sandbox internal DOM instead of the browser's DOM\@. These modifications must 
be committed to the browser's DOM to become visible (line~\ref{line:appendix/motivation/sbx2/dom}).

Alternatively, \SBX\ allows us to grant access to
the browser's DOM by white listing the \lstinline{window} and
\lstinline{document} objects. However, white listing can only expose
entire objects and cannot restrict access to certain parts of the
document model. 

% _   _    _             _____                          _   _             
%| | | |__(_)_ _  __ _  |_   _| _ __ _ _ _  ___ __ _ __| |_(_)___ _ _  ___
%| |_| (_-< | ' \/ _` |   | || '_/ _` | ' \(_-</ _` / _|  _| / _ \ ' \(_-<
% \___//__/_|_||_\__, |   |_||_| \__,_|_||_/__/\__,_\__|\__|_\___/_||_/__/
%                |___/                                                    

\subsection{Using transactions}
\label{sec:appendix/motivation/transactions}

\begin{figure}[tp]
  \centering
\begin{lstlisting}[name=appendix/motivation, language=HTML5]
<!DOCTYPE html>
  <html lang="en">
  <head>
    <!-- ... -->
    <!-- Checks for conflicts with Datejs -->
    <script type="text/javascript">
      sbx2.applyRule(
        new Rule.Commit(Date, function(sbx, effect) {
          return !sbx.inConflictWith(sbx2, Date);
        });*'\label{line:appendix/motivation/sbx2/rule3}'*
      sbx2.applyRule(
      new Rule.RollbackOn(Date, function(sbx, effect) {
          return sbx.inConflictWith(sbx2, Date);
        });*'\label{line:appendix/motivation/sbx2/rule4}'*
    </script>
    </head>
  <body">
    <!-- ... -->
  </body>
</html>
\end{lstlisting}
\caption{Checking for conflicts. The HTML code first checks for conflicts
  between \emph{Datejs} and \emph{jQuery} before it commits the modification of
  the library or rolls back.}
  \label{fig:appendix/overview/example2}
\end{figure}

For wrapped objects, \SBX\ supports a commit/rollback mechanism. In the first
examples (Figure~\ref{fig:appendix/motivation/sandbox}), we prevent the application state from
unintended modification when loading untrusted code and commit only intended ones.

However, \emph{Datejs} and \emph{jquery.formatDateTime.js} might both modify
JavaScript's native \lstinline{Date} object. To avoid undesired overwrites,
\SBX\ allows us to inspect the effects of both libraries and to check for
conflicts before committing to \lstinline{Date}.
The predicate in Line~\ref{line:appendix/motivation/sbx2/rule3}
checks for \emph{conflicts}, which arise in the comparison between different
sandboxes. A conflict is flagged if at least one sandbox modifies a value that is accessed by the other
sandbox later on\footnote{ 
We consider only \emph{Read-After-Write} and \emph{Write-After-Write} conflicts.
\emph{Write-after-Read} conflicts are not handled because the hazard represents
a problem that only occurs in concurrent executions.}.

Furthermore, we prescribe that in case of conflicts the methods from
\lstinline{Datejs} should be used. To this end, a second rule discards the
modifications on \lstinline{Date} from the second sandbox when detecting
conflicts. The \emph{rollback} operation undoes an existing manipulation and
returns to its previous configuration.
Such a partial rollback does not result in an
inconsistent state as we do not delete objects and the references
inside the sandbox remain unchanged.

\subsection{Full HTML Example}
\label{sec:appendix/appendix/motivation/full}

\begin{figure*}[p]
  \centering
\begin{lstlisting}[name=appendix/motivation, language=HTML5]
<!DOCTYPE html>
  <html lang="en">
  <head>
    <!-- DecentJS code-->
    <script src="decent.js"></script>
    <!-- Runs Datejs in a fresh sandbox. -->
    <script type="text/javascript">
      var sbx = new Sandbox(this, Sandbox.DEFAULT);
      sbx.request("datejs.js");
      sbx.applyRule(new Rule.CommitOn(Date, function(sbx, effect) {
          return (effect instanceof Effect.Set) && !(effect.name in Date);
      }));
    </script>
    <!-- Runs jQuery in a fresh sandbox. -->
    <script type="text/javascript">
      var sbx2 = new Sandbox(this, Sandbox.WEB);
      sbx2.initialize("template.html");
      sbx2.applyRule(new Rule.Commit(this, "jQuery"));
      sbx2.applyRule(new Rule.Commit(this, "*'\$'*"));
    </script>
    <!-- Checks for conflicts with Datejs -->
    <script type="text/javascript">
      sbx2.applyRule(new Rule.Commit(Date, function(sbx, effect) {
          return !sbx.inConflictWith(sbx2, Date);
        });
      sbx2.applyRule(new Rule.RollbackOn(Date, function(sbx, effect) {
          return sbx.inConflictWith(sbx2, Date);
        });
    </script>
  </head>
  <body>
    <!-- Body of the page -->
    <h1 id="headline">Headline</h1>
    <script type="text/javascript">
      window.*'\$'*("*'\#'*headline").text("Changed Headline");
      document.getElementById("headline").innerHTML=sbx2.dom.document.getElementById("headline").innerHTML;
    </script>
  </body>
</html>
\end{lstlisting}
\caption{Execution of library code in a sandbox (cf.\ Section~\ref{sec:appendix/motivation}
in the paper).
The first \lstinline[language=HTML5]{<script>} tag loads the sandbox implementation. 
The second \lstinline[language=HTML5]{<script>} tag instantiates a new sandbox
\lstinline{sbx} and loads and executes \emph{Datejs} inside the sandbox. Later
it commits intended effects to the native \lstinline{Date} object.
The third \lstinline[language=HTML5]{<script>} tag instantiates sandbox
\lstinline{sbx2} and initializes the sandbox with a predefined HTML template
(see Figure~\ref{fig:appendix/motivation/template} in the paper). Later it commits intended
modifications to the application state.
The last \lstinline[language=HTML5]{<script>} tag checks for conflicts between
\emph{Datejs} and \emph{jQuery} before it commits further modification on
\lstinline{Date} or rolls back.
The \lstinline[language=HTML5]{<script>} tag included in the body performs a
modification of the sandbox internal DOM and copies the changes to the global
DOM,
}\label{fig:appendix/sandbox/full}
\end{figure*}

Figure~\ref{fig:appendix/sandbox/full} shows the full html code from the example
in Section~\ref{sec:appendix/motivation}.

\subsection{Effects Lists}
\label{sec:appendix/appendix/motivation/effects}

This sections shows the resulting effect logs recorded by the sandboxes in
Section~\ref{sec:appendix/motivation}. See Appendix~\ref{sec:example} for a detailed
explanation of the output.

\subsection{Effects of \lstinline{sbx}}

\subsubsection{All Read Effects on \lstinline{this}}

\begin{lstlisting}[name=appendix/motivation]
sbx.readeffectOn(this).forEach(function(e) {
  print(e);
});
\end{lstlisting}

\begin{lstlisting}[name=appendix/motivation]
(#0) has [name=Date]
(#0) get [name=Date]
(#0) has [name=Number]
(#0) get [name=Number]
(#0) has [name=RegExp]
(#0) get [name=RegExp]
(#0) has [name=Array]
(#0) get [name=Array]
\end{lstlisting}

\subsubsection{All Write Effects on \lstinline{this}}

\begin{lstlisting}[name=appendix/motivation]
sbx.writeeffectOn(this).forEach(function(e) {
  print(e);
});
\end{lstlisting}

\begin{lstlisting}[name=appendix/motivation]
none
\end{lstlisting}

\subsubsection{All Read Effects on \lstinline{Date}}

\begin{lstlisting}[name=appendix/motivation]
sbx.readeffectOn(Date).forEach(function(e) {
  print(e);
});
\end{lstlisting}

\begin{lstlisting}[name=appendix/motivation]
(#1) get [name=prototype]
(#1) get [name=Parsing]
(#1) get [name=Grammar]
(#1) get [name=Translator]
(#1) get [name=CultureInfo]
(#1) get [name=parse]
\end{lstlisting}

\subsubsection{All Write Effects on \lstinline{Date}}

\begin{lstlisting}[name=appendix/motivation]
sbx.writeeffectOn(Date).forEach(function(e) {
  print(e);
});
\end{lstlisting}

\begin{lstlisting}[name=appendix/motivation]
(#1) set [name=CultureInfo]
(#1) set [name=getMonthNumberFromName]
(#1) set [name=getDayNumberFromName]
(#1) set [name=isLeapYear]
(#1) set [name=getDaysInMonth]
(#1) set [name=getTimezoneOffset]
(#1) set [name=getTimezoneAbbreviation]
(#1) set [name=_validate]
(#1) set [name=validateMillisecond]
(#1) set [name=validateSecond]
(#1) set [name=validateMinute]
(#1) set [name=validateHour]
(#1) set [name=validateDay]
(#1) set [name=validateMonth]
(#1) set [name=validateYear]
(#1) set [name=now]
(#1) set [name=today]
(#1) set [name=Parsing]
(#1) set [name=Grammar]
(#1) set [name=Translator]
(#1) set [name=_parse]
(#1) set [name=parse]
(#1) set [name=getParseFunction]
(#1) set [name=parseExact]
\end{lstlisting}

\subsubsection{All Read Effects on \lstinline{Date.prototype}}

\begin{lstlisting}[name=appendix/motivation]
sbx.readeffectOn(Date.prototype).forEach(function(e) {
  print(e);
});
\end{lstlisting}

\begin{lstlisting}[name=appendix/motivation]
(#2) get [name=toString]
\end{lstlisting}

\subsubsection{All Write Effects on \lstinline{Date.prototype}}

\begin{lstlisting}[name=appendix/motivation]
sbx.writeeffectOn(Date.prototype).forEach(function(e) {
  print(e);
});
\end{lstlisting}

\begin{lstlisting}[name=appendix/motivation]
(#2) set [name=clone]
(#2) set [name=compareTo]
(#2) set [name=equals]
(#2) set [name=between]
(#2) set [name=addMilliseconds]
(#2) set [name=addSeconds]
(#2) set [name=addMinutes]
(#2) set [name=addHours]
(#2) set [name=addDays]
(#2) set [name=addWeeks]
(#2) set [name=addMonths]
(#2) set [name=addYears]
(#2) set [name=add]
(#2) set [name=set]
(#2) set [name=clearTime]
(#2) set [name=isLeapYear]
(#2) set [name=isWeekday]
(#2) set [name=getDaysInMonth]
(#2) set [name=moveToFirstDayOfMonth]
(#2) set [name=moveToLastDayOfMonth]
(#2) set [name=moveToDayOfWeek]
(#2) set [name=moveToMonth]
(#2) set [name=getDayOfYear]
(#2) set [name=getWeekOfYear]
(#2) set [name=isDST]
(#2) set [name=getTimezone]
(#2) set [name=setTimezoneOffset]
(#2) set [name=setTimezone]
(#2) set [name=getUTCOffset]
(#2) set [name=getDayName]
(#2) set [name=getMonthName]
(#2) set [name=_toString]
(#2) set [name=toString]
(#2) set [name=_orient]
(#2) set [name=next]
(#2) set [name=previous]
(#2) set [name=prev]
(#2) set [name=last]
(#2) set [name=_is]
(#2) set [name=is]
(#2) set [name=sun]
(#2) set [name=sunday]
(#2) set [name=mon]
(#2) set [name=monday]
(#2) set [name=tue]
(#2) set [name=tuesday]
(#2) set [name=wed]
(#2) set [name=wednesday]
(#2) set [name=thu]
(#2) set [name=thursday]
(#2) set [name=fri]
(#2) set [name=friday]
(#2) set [name=sat]
(#2) set [name=saturday]
(#2) set [name=jan]
(#2) set [name=january]
(#2) set [name=feb]
(#2) set [name=february]
(#2) set [name=mar]
(#2) set [name=march]
(#2) set [name=apr]
(#2) set [name=april]
(#2) set [name=may]
(#2) set [name=jun]
(#2) set [name=june]
(#2) set [name=jul]
(#2) set [name=july]
(#2) set [name=aug]
(#2) set [name=august]
(#2) set [name=sep]
(#2) set [name=september]
(#2) set [name=oct]
(#2) set [name=october]
(#2) set [name=nov]
(#2) set [name=november]
(#2) set [name=dec]
(#2) set [name=december]
(#2) set [name=milliseconds]
(#2) set [name=millisecond]
(#2) set [name=seconds]
(#2) set [name=second]
(#2) set [name=minutes]
(#2) set [name=minute]
(#2) set [name=hours]
(#2) set [name=hour]
(#2) set [name=days]
(#2) set [name=day]
(#2) set [name=weeks]
(#2) set [name=week]
(#2) set [name=months]
(#2) set [name=month]
(#2) set [name=years]
(#2) set [name=year]
(#2) set [name=toJSONString]
(#2) set [name=toShortDateString]
(#2) set [name=toLongDateString]
(#2) set [name=toShortTimeString]
(#2) set [name=toLongTimeString]
(#2) set [name=getOrdinal]
\end{lstlisting}

\subsection{Effects of \lstinline{sbx2}}

\subsubsection{All Read Effects on \lstinline{this}}

\begin{lstlisting}[name=appendix/motivation]
sbx2.readeffectOn(this).forEach(function(e) {
  print(e);
});
\end{lstlisting}

\begin{lstlisting}[name=appendix/motivation]
(#0) has [name=window]
(#0) get [name=window]
(#0) has [name=module]
(#0) get [name=module]
(#0) has [name=Math]
(#0) get [name=Math]
(#0) has [name=Array]
(#0) get [name=Array]
(#0) has [name=Date]
(#0) get [name=Date]
(#0) has [name=undefined]
(#0) get [name=undefined]
(#0) has [name=Symbol]
(#0) get [name=Symbol]
(#0) has [name=RegExp]
(#0) get [name=RegExp]
(#0) has [name=String]
(#0) get [name=String]
(#0) has [name=define]
(#0) get [name=define]
\end{lstlisting}

\subsubsection{All Write Effects on \lstinline{this}}

\begin{lstlisting}[name=appendix/motivation]
sbxs.writeeffectOn(this).forEach(function(e) {
  print(e);
});
\end{lstlisting}

\begin{lstlisting}[name=appendix/motivation]
none
\end{lstlisting}

\subsubsection{All Read Effects on \lstinline{window}}

\begin{lstlisting}[name=appendix/motivation]
sbx2.readeffectOn(window).forEach(function(e) {
  print(e);
});
\end{lstlisting}

\begin{lstlisting}[name=appendix/motivation]
(#1) get [name=window]
(#1) getOwnPropertyDescriptor [name=window]
(#1) getOwnPropertyDescriptor [name=module]
(#1) get [name=document]
(#1) getOwnPropertyDescriptor [name=Math]
(#1) getOwnPropertyDescriptor [name=Array]
(#1) getOwnPropertyDescriptor [name=Date]
(#1) getOwnPropertyDescriptor [name=undefined]
(#1) getOwnPropertyDescriptor [name=Symbol]
(#1) getOwnPropertyDescriptor [name=RegExp]
(#1) get [name=top]
(#1) get [name=setTimeout]
(#1) has [name=onfocusin]
(#1) get [name=location]
(#1) getOwnPropertyDescriptor [name=String]
(#1) get [name=XMLHttpRequest]
(#1) getOwnPropertyDescriptor [name=define]
(#1) get [name=jQuery]
(#1) get [name=$]
\end{lstlisting}

\subsubsection{All Write Effects on \lstinline{window}}

\begin{lstlisting}[name=appendix/motivation]
sbx.writeeffectOn(window).forEach(function(e) {
  print(e);
});
\end{lstlisting}

\begin{lstlisting}[name=appendix/motivation]
(#1) set [name=$]
(#1) set [name=jQuery]
\end{lstlisting}

\cleardoublepage
%                      _ _           _   _             
%    /\               | (_)         | | (_)            
%   /  \   _ __  _ __ | |_  ___ __ _| |_ _  ___  _ __  
%  / /\ \ | '_ \| '_ \| | |/ __/ _` | __| |/ _ \| '_ \ 
% / ____ \| |_) | |_) | | | (_| (_| | |_| | (_) | | | |
%/_/    \_\ .__/| .__/|_|_|\___\__,_|\__|_|\___/|_| |_|
%         | |   | |                                    
%         |_|   |_|                                    
%  _____                           _           
% / ____|                         (_)          
%| (___   ___ ___ _ __   __ _ _ __ _  ___  ___ 
% \___ \ / __/ _ \ '_ \ / _` | '__| |/ _ \/ __|
% ____) | (_|  __/ | | | (_| | |  | | (_) \__ \
%|_____/ \___\___|_| |_|\__,_|_|  |_|\___/|___/

\section{Application Scenarios}
\label{sec:appendix/application}

This section considers some example scenarios that use the implemented system.
All examples are drawn from other projects and use this work's sandboxing
mechanism to guarantee noninterference.

% _____             _      _ ___ 
%|_   _| _ ___ __ _| |_ _ | / __|
%  | || '_/ -_) _` |  _| || \__ \
%  |_||_| \___\__,_|\__|\__/|___/

\subsection{TreatJS}
\label{sec:appendix/application/treatjs}

\TJS\ \cite{KeilThiemann2015-treatjs} is a higher-order contract system for
JavaScript which enforces contracts by run-time monitoring. \TJS\ is implemented
as a library so that all aspects of a contract can be specified using the full
JavaScript language.

For example, the base contract \lstinline|typeNumber| checks its argument to be
a number.
\begin{lstlisting}[name=treatjs]
var |typeNumber| = Contract.Base(function (arg) {*'\label{line:type-number}'*
  return (typeof arg) === 'number';
});
\end{lstlisting}
Asserting a base contracts to a value causes the predicate to be checked
by applying the predicate to the value. In JavaScript, any function can be used
as any return value can be converted to boolean\footnote{%
  JavaScript programmers speak of \emph{truthy} or \emph{falsy} about values
  that convert to true or false.
}.
\begin{lstlisting}[name=treatjs]
Contract.assert(1, |typeNumber|); // accepted
\end{lstlisting}
\TJS\ relies on the sandbox presented in this work to guarantee that
the execution of contract code does not interfere with the contract
abiding execution of the host program.

As read-only access to objects and functions is safe and useful in many
contracts, \TJS\ facilitates making external references visible inside of the
sandbox.

For example, the \lstinline{isArray} contract below references the global object
\lstinline{Array}.
\begin{lstlisting}[name=treatjs]
var |isArray| = Contract.With({Array:Array}, Contract.Base(function (arg) *'\label{line:isarray}'*{
  return (arg instanceof Array);
}));
\end{lstlisting}
However, \TJS\ forbids all write accesses and traps the unintended write to the
global variable \lstinline{type} in the following code.
\begin{lstlisting}[name=treatjs]
var typeNumberBroken = Contract.Base(function(arg) {
  type = (typeof arg); *'\label{line:novar}'*
  return type === 'number';
});
\end{lstlisting}

% _____             _      _ ___    ___       _ _          
%|_   _| _ ___ __ _| |_ _ | / __|  / _ \ _ _ | (_)_ _  ___ 
%  | || '_/ -_) _` |  _| || \__ \ | (_) | ' \| | | ' \/ -_)
%  |_||_| \___\__,_|\__|\__/|___/  \___/|_||_|_|_|_||_\___|

\subsection{TreatJS Online}
\label{sec:appendix/application/treatjsonline}

\TJS-Online\footnote{%
  \url{http://www2.informatik.uni-freiburg.de/~keilr/treatjs/}
} \cite{KeilThiemann:TreatJSOnline} is a web frontend for experimentation with
the \TJS\ contract system~\cite{KeilThiemann2015-treatjs}. It enables the user
to enter code fragments that run in combination with the \TJS\ code. All aspects
of \TJS\ are accessible to the user code. However, the user code should neither
be able to compromise the contract system nor the website's functioning by
writing to the browser's \lstinline{document} or \lstinline{window} objects.
Without any precaution, a code snippet like
\begin{lstlisting}[name=treatjsonline]
Contract.assert = function(arg) {
  return arg;
}
\end{lstlisting}
could change the \lstinline{Contract} objects to influence subsequent
executions,

To avoid these issues, the website creates a fresh sandbox environment, builds a
function closure with the user's input, and executes the user code in the
sandbox. The sandbox grants read-only access to the \TJS\ API and to
JavaScript's built-in objects like \lstinline{Object}, \lstinline{Function},
\lstinline{Array}, and so on, but it does not provide access to browser objects
like \lstinline{document} and \lstinline{window}. Further, each new invocation
reverts the sandbox to its initial state.

%  ___  _                              ___             _        
% / _ \| |__ ___ ___ _ ___ _____ _ _  | _ \_ _ _____ _(_)___ ___
%| (_) | '_ (_-</ -_) '_\ V / -_) '_| |  _/ '_/ _ \ \ / / -_|_-<
% \___/|_.__/__/\___|_|  \_/\___|_|   |_| |_| \___/_\_\_\___/__/

\subsection{Observer Proxies}
\label{sec:appendix/application/observer}

An observer proxy\footnote{%
  This observer proxy in this Subsection should not be confused with the
  \emph{observer proxy} mention in the paper. The observer mentions in
  Section~\ref{sec:sandbox/dom} is a normal proxy implementing a membrane.
} is a restricted version of a JavaScript proxy that cannot change the behavior
of the proxy's target arbitrarily. It implements a projection in that it either
implements the same behavior as the target object or it raises an exception. A
similar feature is provided by Racket's
chaperones~\cite{StricklandTobinHochstadtFindlerFlatt2012}.

\begin{figure}[t]
  \centering
  \begin{lstlisting}[name=example]
  function Observer(target, handler) {
    var sbx = new Sandbox({}, {/* parameters omitted */}); *'\label{line:observersbx}'*
    var controller = { 
      get: function(target, name, receiver) {
        var trap = handler.get;
        var result = trap && sbx.call(trap, target, name, receiver);
        var raw = target[name];
        return observerOf(raw, result) ? result : raw;*'\label{line:return}'*
      }};
      return new Proxy(target, controller);
    }
  \end{lstlisting}
  \caption{Implementation of an observer proxy (excerpt). The \lstinline{get}
  trap evaluates the user specific trap in a sandbox to guarantee
  noninterference. Afterwards it performs the usual operation and compares the
  outcomes of both executions. Other traps can be implemented in the same way.}
  \label{fig:observer}
\end{figure}

Such an {observer} can cause a program to fail more often, but in case it does
not fail it would behave in the same way as if not observer were present.

Figure~\ref{fig:observer} contains the getter part of the JavaScript
implementation of \lstinline{Observer}, the constructor of an observer proxy. It
accepts the same arguments as the constructor of a normal proxy object. It
returns a proxy, but interposes a different handler, \lstinline{controller},
that wraps the execution of all user provided traps in a sandbox.

The controller's \lstinline{get} trap evaluates the user's \lstinline{get} trap
(if one exists) in a sandbox. Next, it performs a normal property access on the
target value to produce the same side effects and to obtain a baseline value to
compare the results. \lstinline{observerOf} checks whether the sandboxed result
is suitably related to the baseline value.

\cleardoublepage
%  _____                            _   _                   __ 
% / ____|                          | | (_)                 / _|
%| (___   ___ _ __ ___   __ _ _ __ | |_ _  ___ ___    ___ | |_ 
% \___ \ / _ \ '_ ` _ \ / _` | '_ \| __| |/ __/ __|  / _ \|  _|
% ____) |  __/ | | | | | (_| | | | | |_| | (__\__ \ | (_) | |  
%|_____/ \___|_| |_| |_|\__,_|_| |_|\__|_|\___|___/  \___/|_|  
%                                                              
%                                                              
%  _____                 _ _               _             
% / ____|               | | |             (_)            
%| (___   __ _ _ __   __| | |__   _____  ___ _ __   __ _ 
% \___ \ / _` | '_ \ / _` | '_ \ / _ \ \/ / | '_ \ / _` |
% ____) | (_| | | | | (_| | |_) | (_) >  <| | | | | (_| |
%|_____/ \__,_|_| |_|\__,_|_.__/ \___/_/\_\_|_| |_|\__, |
%                                                   __/ |
%                                                  |___/ 

\section{Semantics of Sandboxing}
\label{sec:appendix/semantics}

This section first introduces $\lj$, an untyped call-by-value lambda calculus with objects and object proxies that serves as a core calculus for JavaScript, inspired by previous work~\cite{GuhaSaftoiuKrishnamurthi2010,KeilThiemann2015-treatjs-techrep}. It defines its syntax and describes its semantics informally. Later on we extends $\lj$ to a new calculus $\lsbx$, which adds a sandbox to the core calculus.

% ___          _            
%/ __|_  _ _ _| |_ __ ___ __
%\__ \ || | ' \  _/ _` \ \ /
%|___/\_, |_||_\__\__,_/_\_\
%     |__/                  

\subsection{Core Syntax of $\lj$}
\label{sec:appendix/semantics/syntax}

\begin{figure}
  \vspace{-\baselineskip}
  \begin{displaymath}
    \begin{array}{ll@{~}r@{~}l}
      \Label{Constant} &\ni\ljConst\\
      \Label{Variable} &\ni\ljVarx,\ljVary\\
      \Label{Expression} &\ni\ljExpe,\ljExpf,\ljExpg
      &\bbc& \ljConst 
      \mid \ljVar
      \mid \ljOp\ljExpe\ljExpf
      \mid \ljAbs\ljVarx\ljExpe
      \mid \ljApp\ljExpe\ljExpf\\
      &&\mid& \ljNew\ljExpe
      \mid \ljGet\ljExpe\ljExpf
      \mid \ljPut\ljExpe\ljExpf\ljExpg\\
    \end{array}
  \end{displaymath}
  \caption{Syntax of $\lj$.}
  \label{fig:syntax_lj}
\end{figure}

Figure~\ref{fig:syntax_lj} defines the syntax of $\lj$. A $\lj$ expression is either a constant, a variable, an operation on primitive values, a lambds abstraction, an application, a creation of an empty object, a property read, or a property assignment. Variables $\ljVar$, $\ljVary$ are drawn from denumerable sets of symbols and constants $\ljConst$ include JavaScript's primitive values like numbers, strings, booleans, as well as \emph{undefined} and \emph{null}.

The syntax do not make proxies available to the user, but offers an internal method to wrap objects.

% ___                     _   _      ___                 _         
%/ __| ___ _ __  __ _ _ _| |_(_)__  |   \ ___ _ __  __ _(_)_ _  ___
%\__ \/ -_) '  \/ _` | ' \  _| / _| | |) / _ \ '  \/ _` | | ' (_-<
%|___/\___|_|_|_\__,_|_||_\__|_\__| |___/\___/_|_|_\__,_|_|_||_/__/

\subsection{Semantic Domains}
\label{sec:appendix/semantics/domains}

\begin{figure}
  \vspace{-\baselineskip}
  \begin{displaymath}
    \begin{array}{llrl}  
      \Label{Value} &\ni \ljValu,\ljValv,\ljValw &\bbc& \ljConst \mid \ljLoc\\
      \\
      \Label{Closure} &\ni \ljClosure &\bbc& \ljNoClosure \mid (\ljEnv,\ljAbs\ljVarx\ljExpe)\\
      \Label{Dictionary} &\ni \ljDic &\bbc& \emptyset \mid \ljDic[\ljConst\mapsto\ljVal]\\
      \Label{Object} &\ni \ljObj &\bbc& (\ljDic, \ljClosure, \ljVal)\\
      \\
      \Label{Environment} &\ni \ljEnv &\bbc& \emptyset \mid \ljEnv[\ljVar\mapsto\ljVal]\\
      \Label{Store} &\ni \ljStore &\bbc& \emptyset \mid \ljStore[\ljLoc\mapsto\ljObj]\\
    \end{array}
  \end{displaymath}
  \caption{Semantic domains of $\lj$.}
  \label{fig:domains_lj}
\end{figure}

\begin{figure}[]
  \vspace{-\baselineskip}
  \begin{displaymath}
    \begin{array}{ll@{~}r@{~}l}
      \Label{Sandbox} &\ni\sbx
      &\bbc& \ljSbx\ljVar\ljExpe\\

      \\
      \Label{Expression} &\ni\ljExpe,\ljExpf,\ljExpg
      &\bbc& \cdots \mid \sbx \mid \ljFresh\ljExp\\

      \Label{Term} &\ni\ljTerm
      &\bbc& \cdots \mid \ljFresh\sbx \mid \ljWrap\ljVal\\

      \\
      \Label{Object} &\ni\ljObj 
      &\bbc& \cdots \mid (\ljLoc, \ljSLoc, \ljSEnv)\\

      \Label{Values} &\ni\ljValu,\ljValv,\ljValw
      &\bbc& \cdots \mid (\ljSEnv, \sbx)\\

    \end{array}
  \end{displaymath}
  \caption{Extensions of $\lsbx$.}\label{fig:extension_lsbx}
\end{figure}

Figure~\ref{fig:domains_lj} defines the semantic domains of $\lj$.

Its main component is a store that maps a location $\ljLoc$ to an object $\ljObj$, which is a native object (non-proxy object) represented by a triple consisting of a dictionary $\ljDic$, a potential function closure $\ljClosure$, and a value $\ljVal$ acting as prototype. A dictionary $\ljDic$ models the properties of an object. It maps a constant $\ljConst$ to a value $\ljVal$. An object may be a function in which case its closure consists of a lambda expression $\ljAbs\ljVarx\ljExpe$ and an environment $\ljEnv$ that binds the free variables. It maps a variable $\ljVar$ to a value $\ljVal$. A non-function object is indicated by $\ljNoClosure$ in this place.

A value $\ljVal$ is either a constant $\ljConst$ or a location $\ljLoc$.

% ___          _           _   _          
%| __|_ ____ _| |_  _ __ _| |_(_)___ _ _  
%| _|\ V / _` | | || / _` |  _| / _ \ ' \ 
%|___|\_/\__,_|_|\_,_\__,_|\__|_\___/_||_|

\subsection{Evaluation of $\lj$}
\label{sec:appendix/semantics/evaluation}

\begin{figure}
  \vspace{-\baselineskip}
  \begin{displaymath}
    \begin{array}{ll@{~}r@{~}l}
      \Label{Term} &\ni \ljTerm &\bbc& \ljExp\\ 
      &&\mid& \ljOp\ljValv\ljExpf
      \mid \ljOp\ljVal\ljValw
      \mid \ljApp\ljLoc\ljExpf
      \mid \ljApp\ljLoc\ljValv
      \mid \ljNew\ljValv\\
      &&\mid& \ljGet\ljLoc\ljExpf
      \mid \ljGet\ljLoc\ljConst
      \mid \ljPut\ljLoc\ljExpf\ljExpg
      \mid \ljPut\ljLoc\ljConst\ljExpg
      \mid \ljPut\ljLoc\ljConst\ljValw
    \end{array}
  \end{displaymath}
  \caption{Intermediate terms of $\lj$.}
  \label{fig:terms_lj}
\end{figure}

\begin{figure*}[p]
  \centering
  \begin{mathpar}
    \inferrule[\RuleLjConst]
    {%
    }
    {%
      \ljEnv\entails\ljStore,\ljConst\eval\ljStore,\ljConst%
    }\and
    \inferrule[\RuleLjVar]
    {%
    }
    {%
      \ljEnv\entails\ljStore,\ljVar\eval\ljStore,\ljEnv(\ljVar)%
    }\and
    \inferrule[\RuleLjOpE]
    {%
      \ljEnv\entails\ljStore,\ljExpe\eval\ljStore',\ljValv\\\\
      \ljEnv\entails\ljStore',\ljOp\ljValv\ljExpf\eval\ljStore'',\ljValw%
    }
    {%
      \ljEnv\entails\ljStore,\ljOp\ljExpe\ljExpf\eval\ljStore'',\ljValw%
    }\and
    \inferrule[\RuleLjOpF]
    {%
      \ljEnv\entails\ljStore,\ljExpf\eval\ljStore',\ljValu\\\\
      \ljEnv\entails\ljStore',\ljOp\ljVal\ljValu\eval\ljStore'',\ljValw%
    }
    {%
      \ljEnv\entails\ljStore,\ljOp\ljVal\ljExpf\eval\ljStore'',\ljValw%
    }\and
    \inferrule[\RuleLjOp]
    {%
      \ljValw=\ljOp\ljVal\ljValu%
    }
    {%
      \ljEnv\entails\ljStore,\ljOp\ljVal\ljValu\eval\ljStore,\ljValw%
    }\and
    \inferrule[\RuleLjAbs]
    {%
      \ljLoc\notin\dom{\ljStore}\\
      \ljStore'=\ljStore[\ljLoc\mapsto(\emptyset, (\ljEnv,\ljAbs\ljVarx\ljExpe),\ljNull)]
    }
    {%
      \ljEnv\entails\ljStore,\ljAbs\ljVarx\ljExpe\eval\ljStore',\ljLoc%
    }\and
    \inferrule[\RuleLjAppE]
    {%
      \ljEnv\entails\ljStore,\ljExp\eval\ljStore',\ljLoc\\\\
      \ljEnv\entails\ljStore',\ljApp\ljLoc\ljExpf\eval\ljStore'',\ljValw%
    }
    {%
      \ljEnv\entails\ljStore,\ljApp\ljExp\ljExpf\eval\ljStore'',\ljValw%
    }\and
    \inferrule[\RuleLjAppF]
    {%
      \ljEnv\entails\ljStore,\ljExpf\eval\ljStore',\ljValv\\\\
      \ljEnv\entails\ljStore',\ljApp\ljLoc\ljValv\eval\ljStore'',\ljValw%
    }
    {%
      \ljEnv\entails\ljStore,\ljApp\ljLoc\ljExpf\eval\ljStore'',\ljValw%
    }\and
    \inferrule[\RuleLjApp]
    {%
      (\ljDic, (\ljEnv',\ljAbs\ljVarx\ljExpe), \ljValu)=\ljStore(\ljLoc)\\
      \ljEnv'[\ljVar\mapsto\ljVal]\entails\ljStore,\ljExp\eval\ljStore',\ljValw%
    }
    {%
      \ljEnv\entails\ljStore,\ljApp\ljLoc\ljVal\eval\ljStore',\ljValw%
    }\and
    \inferrule[\RuleLjNewE]
    {%
      \ljEnv\entails\ljStore,\ljExp\eval\ljStore',\ljValv\\\\
      \ljEnv\entails\ljStore',\ljNew\ljValv\eval\ljStore'',\ljValw%
    }
    {%
      \ljEnv\entails\ljStore,\ljNew\ljExp\eval\ljStore'',\ljValw%
    }\and
    \inferrule[\RuleLjNew]
    {%
      \ljLoc\notin\dom{\ljStore}\\
      \ljStore'=\ljStore[\ljLoc\mapsto(\emptyset,\ljNoClosure,\ljVal)]
    }
    {%
      \ljEnv\entails\ljStore,\ljNew\ljVal\eval\ljStore',\ljLoc%
    }\and
    \inferrule[\RuleLjGetE]
    {%
      \ljEnv\entails\ljStore,\ljExp\eval\ljStore',\ljLoc\\\\
      \ljEnv\entails\ljStore',\ljGet\ljLoc\ljExpf\eval\ljStore'',\ljValw%
    }
    {%
      \ljEnv\entails\ljStore,\ljGet\ljExp\ljExpf\eval\ljStore'',\ljValw%
    }\and
    \inferrule[\RuleLjGetF]
    {%
      \ljEnv\entails\ljStore,\ljExpf\eval\ljStore',\ljConst\\\\
      \ljEnv\entails\ljStore',\ljGet\ljLoc\ljConst\eval\ljStore'',\ljValw%
    }
    {%
      \ljEnv\entails\ljStore,\ljGet\ljLoc\ljExpf\eval\ljStore'',\ljValw%
    }\and
    \inferrule[\RuleLjGet]
    {%
      (\ljDic,\ljClosure,\ljVal)=\ljStore(\ljLoc)\\
      \ljConst\in\dom\ljDic%
    }
    {%
      \ljEnv\entails\ljStore,\ljGet\ljLoc\ljConst\eval\ljStore,\ljDic(\ljConst)
    }\and
    \inferrule[\RuleLjGetProto]
    {%
      (\ljDic,\ljClosure,\ljLoc') = \ljStore(\ljLoc)\\
      \ljConst\notin\dom\ljDic\\\\
      \ljEnv\entails\ljStore,\ljGet{\ljLoc'}\ljConst\eval\ljStore,\ljValv%
    }
    {%
      \ljEnv\entails\ljStore,\ljGet\ljLoc\ljConst\eval\ljStore,\ljValv%
    }\and
    \inferrule[\RuleLjGetUndef]
    {%
      (\ljDic,\ljClosure,\ljConst')=\ljStore(\ljLoc)\\
      \ljConst\notin\dom\ljDic%
    }
    {%
      \ljEnv\entails\ljStore,\ljGet\ljLoc\ljConst\eval\ljStore,\ljUndefined%
    }\and   
    \inferrule[\RuleLjPutE]
    {%
      \ljEnv\entails\ljStore,\ljExp\eval\ljStore',\ljLoc\\\\
      \ljEnv\entails\ljStore',\ljPut\ljLoc\ljExpf\ljExpg\eval\ljStore'',\ljValw%
    }
    {%
      \ljEnv\entails\ljStore,\ljPut\ljExp\ljExpf\ljExpg\eval\ljStore'',\ljValw%
    }\and
    \inferrule[\RuleLjPutF]
    {%
      \ljEnv\entails\ljStore,\ljExpf\eval\ljStore',\ljConst\\\\
      \ljEnv\entails\ljStore',\ljPut\ljLoc\ljConst\ljExpg\eval\ljStore'',\ljValw%
    }
    {%
      \ljEnv\entails\ljStore,\ljLoc[\ljExpf]\!=\!\ljExpg\eval\ljStore'',\ljValw%
    }\and
    \inferrule[\RuleLjPutG]
    {%
      \ljEnv\entails\ljStore,\ljExpg\eval\ljStore',\ljValv\\\\
      \ljEnv\entails\ljStore',\ljPut\ljLoc\ljConst\ljValv\eval\ljStore'',\ljValw%
    }
    {%
      \ljEnv\entails\ljStore,\ljPut\ljLoc\ljConst\ljExpg\eval\ljStore'',\ljValw%
    }\and
    \inferrule[\RuleLjPut]
    {%
      (\ljDic,\ljClosure,\ljValu)=\ljStore(\ljLoc)\\
      \ljStore'=\ljStore[\ljLoc\mapsto(\ljDic[\ljConst\mapsto\ljVal],\ljClosure,\ljValu)]
    }
    {%
      \ljEnv\entails\ljStore,\ljPut\ljLoc\ljConst\ljVal\eval\ljStore',\ljVal%
    }
  \end{mathpar}
  \caption{Inference rules for intermediate terms of $\lj$.}\label{fig:lj}
\end{figure*}

A pretty-big-step semantics~\cite{Chargueraud2013} introduces intermediate terms to model partially evaluated expressions (Figure~\ref{fig:terms_lj}). An intermediate term is thus an expression where zero or more top-level subexpressions are replaced by their outcomes.

The evaluation judgment is similar to a standard big-step evaluation judgment except that its input ranges over intermediate terms: It states that evaluation of term $\ljTerm$ with initial store $\ljStore$, and environment $\ljEnv$ results in a final store $\ljStore'$ and value $\ljValv$.
\begin{displaymath}
  \ljEnv \entails \langle \ljStore,\ljTerm \rangle \eval \langle
  \ljStore',\ljValv \rangle
\end{displaymath}
Figure~\ref{fig:lj} defines the standard evaluation rules for expressions $\ljExp$ in $\lj$.
The inference rules for expressions $\ljExp$ are mostly standard. Each rule for a composite expression evaluates exactly one subexpression and then recursively invokes the evaluation judgment to continue. Once all subexpressions are evaluated, the respective rule performs the desired operation.

% ___               _ _             _           
%/ __| __ _ _ _  __| | |__  _____ _(_)_ _  __ _ 
%\__ \/ _` | ' \/ _` | '_ \/ _ \ \ / | ' \/ _` |
%|___/\__,_|_||_\__,_|_.__/\___/_\_\_|_||_\__, |
%                                         |___/ 

\subsection{Sandboxing of $\lj$}
\label{sec:appendix/semantics/extension}

This section extends the base calculus $\lj$ to a calculus $\lsbx$ which adds sandboxing of function expressions. The calculus describes only the core features that illustrates the principles of our sandbox. Further features of the application level can be implemented in top of the calculus.

Figure~\ref{fig:extension_lsbx} defines the syntax and semantics of $\lsbx$ as an extension of $\lj$. Expressions now contain a sandbox abstraction $\sbx$ and a sandbox construction $\ljFresh\ljExp$ that instantiates a fresh sandbox.

Terms are extended with a $\ljFresh\sbx$ term. A new internal $\ljWrap\ljVal$ term, which did not occour in source programs, wraps a value in a sandbox environment.

Objects now contain object proxies. A proxy object is a single location controlled by a proxy handler that mediates the access to the target location. For simplification, we represent handler objects by there meta-data. So, each handler is an sandbox handler that enforces write-protection (viz.\ by an \emph{secure} location $\ljSLoc$ that acts as an shadow object for the proxies target object $\ljLoc$ and a single \emph{secure} environment $\ljSEnv$).

For clarity, we write $\ljSVal$, $\ljSValu$, $\ljSValw$ for wrapped values that are imported into a sandbox, $\ljSEnv$ for a sandbox environment that only contains wrapped values, and $\ljSLoc$ for locations of proxies and shadow objects.

Consequently, values are extended with sandboxes which represents an sandbox expression wrapped in a sandbox environment that is to be executed when the value is used in an application.

\subsubsection{Evaluation of $\lsbx$}
\label{sec:appendix/semantics/sandbox}

\begin{figure*}
  \centering
  \begin{mathpar}
    \inferrule[\RuleSbxFreshE]
    {%
      \ljEnv\entails\ljStore,\ljExpe\eval\ljStore',\ljSbx\ljVar\ljExpf\\\\
      \ljEnv\entails\ljStore',\ljFresh{\ljSbx\ljVar\ljExpf} \eval\ljStore'',\ljValv\\
    }
    {%
      \ljEnv\entails\ljStore,\ljFresh\ljExpe\eval\ljStore'',\ljValv%
    }\and
    \inferrule[\RuleSbxFresh]
    {%
      \emptyset\entails\ljStore,\ljSbx\ljVar\ljExpe%
      \eval\ljStore, (\ljSEnv,\ljSbx\ljVar\ljExpe)
    }
    {%
      \ljEnv\entails\ljStore,\ljFresh{\ljSbx\ljVar\ljExpe}%
      \eval\ljStore, (\ljSEnv,\ljSbx\ljVar\ljExpe)
    }\and
    \inferrule[\RuleSbxAbs]
    {}
    {%
      \ljSEnv\entails\ljStore,\ljSbx\ljVar\ljExpe%
      \eval\ljStore, (\ljSEnv,\ljSbx\ljVar\ljExpe)
    }\and
    \inferrule[\RuleSbxAbb]
    {%
      \ljSEnv\entails\ljStore,\ljWrap\ljValv\eval\ljStore',\ljSValv\\\\
      \ljSEnv[\ljVar\mapsto\ljSValv] \entails\ljStore',\ljExpe%
      \eval\ljStore'',\ljSValw\\
    }
    {%
      \ljEnv\entails%
      \ljStore,\ljApp{(\ljSEnv,\ljSbx\ljVar\ljExpe)}{\ljValv}%
      \eval\ljStore'',\ljSValw%
    }
  \end{mathpar}
  \caption{Sandbox abstraction and application rules of $\lsbx$.}\label{fig:sandbox}
\end{figure*}

Figure~\ref{fig:sandbox} contains its inference rules for sandbox abstraction and sandbox application of $\lsbx$. The formalization employs pretty-big-step semantics~\cite{Chargueraud2013} to model side effects while keeping the number of evaluation rules manageable.

The rule for expression $\ljFresh\ljExp$ (Rule \RefTirName{\RuleSbxFreshE}) evaluates the subexpression and invokes the evaluation judgment to continue. The other rules show the last step in a pretty big step evaluation. Once all subexpressions are evaluated, the respective rule performs the desired operation.

Sandbox execution happens in the context of a secure sandbox environment to preserve noninterference. So a sandbox definition (abstraction) will evaluate to a sandbox closure containing the sandbox expression (the abstraction) together with an empty environment (Rule \RefTirName{\RuleSbxFresh}). Each sandbox executions starts from a fresh environment. This guarantees that not unwrapped values are reachable by the sandbox.

Sandbox abstraction (Rule \RefTirName{\RuleSbxAbs}) proceeds only on secure environments, which is either an empty set or an environment that contains only secure (wrapped) values.

Sandbox execution (Rule \RefTirName{\RuleSbxAbb}) applies after the first expression evaluates to a sandbox closure and the second expression evaluates to a value. It wraps the given value and triggers the evaluation of expressions $\ljExpe$ in the sandbox environment $\ljSEnv$ after binding the wrapped value $\ljSValv$. Value $\ljSValv$ acts as the global object of the sandbox. It can be used to make values visible inside ob the sandbox.

\subsubsection{Sandbox Encapsulation}
\label{sec:appendix/semantics/sandbox/wrap}

\begin{figure*}
  \centering
  \begin{mathpar}
    \inferrule[\RuleWrapConst] 
    {}
    {%
      \ljSEnv\entails\ljStore,\ljWrap\ljConst%
      \eval\ljStore,\ljConst%
    }\and
    \inferrule[\RuleWrapSandbox] 
    {}
    {%
      \ljSEnv\entails\ljStore,\ljWrap{(\ljSEnv',\sbx)}\eval\ljStore, (\ljSEnv',\sbx)
    }\and
    \inferrule[\RuleWrapNonProxy]
    {% 
      \not\exists\ljLoc'\in\dom{\ljStore}:~(\ljLoc,\ljSLoc,\ljSEnv)=\ljStore(\ljLoc')\\\\
      \ljSEnv\entails\ljStore,\ljRecomp{\ljLoc}\eval\ljStore',\ljSLoc\\\\
      \ljSLoc'\not\in\dom{\ljStore'}\\
      \ljStore''=\ljStore'[\ljSLoc'\mapsto(\ljLoc,\ljSLoc,\ljSEnv)]
    }
    {%
      \ljSEnv\entails\ljStore,\ljWrap\ljLoc\eval\ljStore'',\ljSLoc'
    }\and
    \inferrule[\RuleWrapExisting]
    {}
    {%
      \ljSEnv\entails\ljStore[\ljSLoc\mapsto(\ljLoc,\ljSLoc',\ljSEnv)],\ljWrap\ljLoc%
      \eval\ljStore,\ljSLoc%
    }\and
    \inferrule[\RuleWrapProxy]
    {}
    {%
      \ljSEnv\entails%
      \ljStore[\ljSLoc\mapsto(\ljLoc,\ljSLoc',\ljSEnv)],\ljWrap{\ljSLoc}%
      \eval\ljStore,\ljSLoc%
    }
  \end{mathpar}
  \caption{Inference rules for sandbox encapsulation.}\label{fig:wrap}
\end{figure*}

The sandbox encapsulation (Figure~\ref{fig:wrap}) distinguishes several cases. A primitive value and a sandbox closure is not wrapped.

To wrap a location that points to a non-proxy object, the location is packed in a fresh proxy along with a fresh shadow object and the current sandbox environment. This packaging ensures that each further access to the wrapped location has to use the current environment.

In case the location is already wrapped by a sandbox proxy or the location of a sandbox proxy gets wrapped then the location to the existing proxy is returned. This rule ensures that an object is wrapped at most once and thus preserves object identity inside the sandbox.

\begin{figure*}
  \centering
  \begin{mathpar}   
    \inferrule[\RuleRecompNonFunction] 
    {%
      \ljSLoc\not\in\dom{\ljStore}\\
      \ljStore'=\ljStore[\ljSLoc\mapsto(\emptyset,\ljNoClosure,\ljNull)]
    }
    {%
      \ljSEnv\entails%
      \ljStore[\ljLoc\mapsto(\ljDic,\ljNoClosure,\ljVal)],\ljRecomp\ljLoc%
      \eval\ljStore,\ljSLoc%
    }\and
    \inferrule[\RuleRecompFunction] 
    {%
      \not\exists\ljSLoc\in\dom{\ljStore}:~(\ljDic, (\ljSEnv,\ljAbs\ljVarx\ljExpe),\ljVal)=\ljStore(\ljSLoc)\\\\
      \ljSLoc\not\in\dom{\ljStore}\\
      \ljStore'=\ljStore[\ljSLoc\mapsto(\emptyset, (\ljSEnv,\ljAbs\ljVarx\ljExpe),\ljNull)]
    }
    {%
      \ljSEnv\entails%
      \ljStore[\ljLoc\mapsto(\ljDic,
      (\ljEnv,\ljAbs\ljVarx\ljExpe),\ljVal)],\ljRecomp\ljLoc%
      \eval\ljStore,\ljSLoc%
    }\and
    \inferrule[\RuleRecompileExisting] 
    {%
    }
    {%
      \ljSEnv\entails%
      \ljStore[\ljSLoc\mapsto(\ljDic, (\ljSEnv,\ljAbs\ljVarx\ljExpe),\ljVal)],\ljRecomp\ljSLoc%
      \eval\ljStore,\ljSLoc%
    }\and
    \inferrule[\RuleRecompileProxy] 
    {%
      \ljSEnv\entails%
      \ljStore[\ljLoc\mapsto(\ljLoc',\ljSLoc,\ljSEnv)],\ljRecomp{\ljLoc'}\eval%
      \ljStore,\ljSLoc%
    }
    {%
      \ljSEnv\entails%
      \ljStore[\ljLoc\mapsto(\ljLoc',\ljSLoc,\ljSEnv)],\ljRecomp\ljLoc\eval%
      \ljStore,\ljSLoc%
    }
  \end{mathpar}
  \caption{Inference rules for object re-compilation.}\label{fig:recompile}
\end{figure*}

The shadow object is build from recompiling (Figure~\ref{fig:recompile}) the target object. A shadow objects is a new empty object that may carry a sandboxed replication of its closure part.

For a non-function object, recompiling returns an empty object that later on acts as a sink for property assignments on the wrapped objects.

For a function object, recompiling extracts the function body from the closure and redefines the body with respect to the current sandbox environment. The new closure is put into a new empty object. This step erases all external bindings of function closure and ensures that the application of a wrapped function happens in the context of the secure sandbox environment.

In case the function is already recompiled, function recompilation returns the existing replication.

\subsubsection{Application, Read, and Assignment}
\label{sec:appendix/semantics/sandbox/trap}

Function application, property read, and property assignment distinguish two cases: either the operation applies directly to a non-proxy object or it applies to a proxy. If the target of the operation is not a proxy object, then the usual rules apply.

\begin{figure*}
  \centering
  \begin{mathpar}   
    \inferrule[\RuleAppSandbox] 
    {%
      \ljSEnv\entails\ljStore,\ljWrap\ljVal\eval\ljStore',\ljSValv\\\\
      \ljEnv\entails\ljStore',\ljSLoc(\ljSVal)\eval\ljStore'',\ljSValw%
    }
    {%
      \ljEnv\entails%
      \ljStore[\ljLoc\mapsto(\ljLoc',\ljSLoc,\ljSEnv)],\ljLoc(\ljVal)\eval%
      \ljStore'',\ljSValw%
    }
    \and
    \inferrule[\RuleGetShadow]
    {%
      \ljConst\in\dom\ljSLoc\\
      \ljEnv\entails\ljStore[\ljLoc\mapsto(\ljLoc',\ljSLoc,\ljSEnv)],\ljGet\ljSLoc\ljConst%
      \eval\ljStore',\ljSValv%
    }
    {%
      \ljEnv\entails%
      \ljStore[\ljLoc\mapsto(\ljLoc',\ljSLoc,\ljSEnv)],\ljGet\ljLoc\ljConst%
      \eval\ljStore',\ljSValv%
    }\and
    \inferrule[\RuleGetSandbox]
    {%
      \ljConst\not\in\dom\ljSLoc\\
      \ljEnv\entails%
      \ljStore[\ljLoc\mapsto(\ljLoc',\ljSLoc,\ljSEnv)],\ljGet{\ljLoc'}\ljConst%
      \eval\ljStore',\ljValv\\\\
      \ljSEnv\entails\ljStore',\ljWrap\ljValv\eval\ljStore'',\ljSValv%
    }
    {%
      \ljEnv\entails%
      \ljStore[\ljLoc\mapsto(\ljLoc',\ljSLoc,\ljSEnv)],\ljGet\ljLoc\ljConst%
      \eval\ljStore'',\ljSValv%
    }\and
    \inferrule[\RulePutSandbox]
    {%
      \ljEnv\entails%
      \ljStore[\ljLoc\mapsto(\ljLoc',\ljSLoc,\ljSEnv)],\ljPut\ljSLoc\ljConst\ljValv%
      \eval\ljStore',\ljValv%
    }
    {%
      \ljEnv\entails%
      \ljStore[\ljLoc\mapsto(\ljLoc',\ljSLoc,\ljSEnv)],\ljPut\ljLoc\ljConst\ljValv%
      \eval\ljStore',\ljValv%
    }
  \end{mathpar}
  \caption{Inference rules for function application, property read, and property
  assignment.}\label{fig:app-get-put}
\end{figure*}

Figure~\ref{fig:app-get-put} contains the inference rules for function application and property access for the non-standard cases.

The application of a wrapped function proceeds by unwrapping the function and evaluating it in the sandbox environment contained in the proxy. The function argument and its result are known to be wrapped in this case.

A property read on a wrapped object has two cases depending on if the accessed property has been written in the sandbox before, or not. The notation $\ljConst\in\dom\ljLoc$ is defined as an shortcut of a dictionary lookup $\ljConst\in\dom\ljDic$ with $\ljStore(\ljLoc)=(\ljDic,\ljClosure,\ljValv)$.

A property read of an affected field reads the property from the shadow location. Otherwise, it continues the operation on the target and wraps the resulting value. An assignment to a wrapped object is continues with the operation on the shadow location $\ljSLoc$.

In JavaScript, write operations do only change properties of the object's dictionary. They do not affect the object's prototype. Therefor, the shadow object did not contain any prototype informations. It acts only a shadow that absorbs write operations.

\cleardoublepage
% _______        _           _           _   _____                 _ _       
%|__   __|      | |         (_)         | | |  __ \               | | |      
%   | | ___  ___| |__  _ __  _  ___ __ _| | | |__) |___  ___ _   _| | |_ ___ 
%   | |/ _ \/ __| '_ \| '_ \| |/ __/ _` | | |  _  // _ \/ __| | | | | __/ __|
%   | |  __/ (__| | | | | | | | (_| (_| | | | | \ \  __/\__ \ |_| | | |_\__ \
%   |_|\___|\___|_| |_|_| |_|_|\___\__,_|_| |_|  \_\___||___/\__,_|_|\__|___/
%                                                                            

\section{Technical Results}
\label{sec:appendix/technicalresults}

As JavaScript is a memory safe programming language, a reference can be seen as
the right ti modify the underlying object. If an expressions body can be shown
not to contain unprotected references to objects, then it cannot modify this
data.

To prove soundness of our sandbox we show termination insensitive
noninterference. It requires to show that the initial store $\ljStore$ of a
sandbox application is \emph{observational equivalent} to the final store
$\ljStore'$, that remains after the application. In detail, the sandbox
application may introduce new objects or even write to shadow objects (only
reachable inside of the sandbox) but every value reachable from the outside
remains unmodified. 

As the calculus in Appendix~\ref{sec:appendix/semantics} did not support variable
updates on environments $\ljEnv$ the only way to make changes persistent is to
modify objects. Thus, proving noninterference relates different stores and looks
for differences in the store with respect to all reachable values.

\subsection{Observational Equivalence on Stores}
\label{sec:equivalence}

First, we introduce an equivalence relation on stores with respect to other
semantic elements.

\begin{definition}\label{def:eq-const}
  Two stores $\ljStore$, $\ljStore'$ are equivalent w.r.t constants $\ljConst$, $\ljConst'$ if the
  constants are equal.
  \begin{mathpar}
    \eqstore{(\ljStore, \ljConst)}{(\ljStore', \ljConst')}
    \Leftrightarrow
    \ljConst=\ljConst
  \end{mathpar}
\end{definition}

\begin{definition}\label{def:eq-loc}
  Two stores $\ljStore$, $\ljStore'$ are equivalent w.r.t locations $\ljLoc$, $\ljLoc'$ if
  they are equivalent on the location's target.
  \begin{mathpar}
    \eqstore{(\ljStore, \ljLoc)}{(\ljStore', \ljLoc')}
    \Leftrightarrow
    \eqstore{(\ljStore, \ljStore(\ljLoc))}{(\ljStore', \ljStore'(\ljLoc'))}
  \end{mathpar}
\end{definition}

\begin{definition}\label{def:eq-obj}
  Two stores $\ljStore$, $\ljStore'$ are equivalent w.r.t non-proxy objects $(\ljDic, \ljClosure,
  \ljVal)$, $(\ljDic', \ljClosure', \ljVal')$ if they are equivalent on the objects's constituents.
  \begin{mathpar}
    \eqstore{(\ljStore, (\ljDic, \ljClosure, \ljVal))}{(\ljStore', (\ljDic', \ljClosure', \ljVal'))}
    \Leftrightarrow\\
    \eqstore{(\ljStore, \ljDic)}{(\ljStore', \ljDic')}
    \wedge
    \eqstore{(\ljStore, \ljClosure)}{(\ljStore', \ljClosure')}
    \wedge
    \eqstore{(\ljStore, \ljVal)}{(\ljStore', \ljVal')}
  \end{mathpar}
\end{definition}

\begin{definition}\label{def:eq-dic}
  Two stores $\ljStore$, $\ljStore'$ are equivalent w.r.t dictionaries $\ljDic$,
  $\ljDic'$ if they are equivalent on the dictionary's content.
  \begin{mathpar}
    \eqstore{(\ljStore, \ljDic)}{(\ljStore', \ljDic')}
    \Leftrightarrow\\
    \dom\ljDic=\dom{\ljDic'}
    \wedge
    \forall\ljConst\in\dom\ljDic.\eqstore{(\ljStore, \ljDic(\ljConst))}{(\ljStore', \ljDic'(\ljConst))}
  \end{mathpar}
\end{definition}

\begin{definition}\label{def:eq-closure}
  Two stores $\ljStore$, $\ljStore'$ are equivalent w.r.t closures $(\ljEnv, \ljAbs\ljVarx\ljExpe)$,
  $(\ljEnv', \ljAbs\ljVarx\ljExpf)$ if the are equivalent on the closure's environment and both
  abstractions are equal.
  \begin{mathpar}
    \eqstore{(\ljStore, (\ljEnv, \ljAbs\ljVarx\ljExpe))}{(\ljStore', (\ljEnv',
    \ljAbs\ljVarx\ljExpf))}
    \Leftrightarrow\\
    \eqstore{(\ljStore, \ljEnv)}{(\ljStore', \ljEnv')} 
    \wedge
    \ljAbs\ljVarx\ljExpe=\ljAbs\ljVarx\ljExpf
  \end{mathpar}
\end{definition}

\begin{definition}\label{def:eq-env}
  Two stores $\ljStore$, $\ljStore'$ are equivalent w.r.t environments $\ljEnv$,
  $\ljEnv'$ if the are equivalent on the environment's content.
  \begin{mathpar}
    \eqstore{(\ljStore, \ljEnv)}{(\ljStore', \ljEnv')} 
    \Leftrightarrow\\
    \dom\ljEnv=\dom{\ljEnv'}
    \wedge
    \forall\ljVar\in\dom\ljEnv.\eqstore{(\ljStore, \ljEnv(\ljVar))}{(\ljStore', \ljEnv(\ljVar))}
  \end{mathpar}
\end{definition}

\begin{definition}\label{def:eq-proxy}
  Two stores $\ljStore$, $\ljStore'$ are equivalent w.r.t proxy objects $(\ljLoc, \ljSLoc,
  \ljSEnv)$, $(\ljLoc', \ljSLoc', \ljSEnv')$ if they are equivalent on the objects's constituents.
  \begin{mathpar}
    \eqstore{(\ljStore, (\ljLoc, \ljSLoc, \ljSEnv))}{(\ljStore', (\ljLoc', \ljSLoc', \ljSEnv'))}
    \Leftrightarrow\\
    \eqstore{(\ljStore, \ljLoc)}{(\ljStore', \ljLoc')}
    \wedge
    \eqstore{(\ljStore, \ljSLoc)}{(\ljStore', \ljSLoc')}
    \wedge
    \eqstore{(\ljStore, \ljSEnv)}{(\ljStore', \ljSEnv')}
  \end{mathpar}
\end{definition}

\begin{definition}\label{def:eq-sbx}
  Two stores $\ljStore$, $\ljStore'$ are equivalent w.r.t sandbox closures $(\ljSEnv,
  \ljSbx\ljVarx\ljExpe)$, $(\ljSEnv', \ljSbx\ljVarx\ljExpf)$ if the are equivalent on the sandbox's
  environment and both abstractions are equal.
  \begin{mathpar}
    \eqstore{(\ljStore, (\ljSEnv,\ljSbx\ljVarx\ljExpe))}{(\ljStore', (\ljSEnv',
    \ljSbx\ljVarx\ljExpf))} 
    \Leftrightarrow\\
    \eqstore{(\ljStore, \ljSEnv)}{(\ljStore', \ljSEnv')} 
    \wedge
    \ljSbx\ljVarx\ljExpe=\ljSbx\ljVarx\ljExpf
  \end{mathpar}
\end{definition}

Now, the observational equivalence for stores can be states as follows.

\begin{definition}\label{def:eq-store}
  Two stores $\ljStore$, $\ljStore'$ are observational equivalent under environment $\ljEnv$ if they
  are equivalent on all values $\ljVal\in\{\ljEnv(\ljVar) \mid \ljVar\in\dom\ljEnv\}$ 
  \begin{mathpar}
    \eqenv{\ljStore}{\ljStore'} \Leftrightarrow \forall\ljVar\in\dom\ljEnv.
    \eqstore{(\ljStore, \ljEnv(\ljVar))}{(\ljStore', \ljEnv(\ljVar))}
  \end{mathpar}
\end{definition}

\begin{lemma}\label{thm:equivalence}
  Suppose that  $\ljEnv_i \entails \langle \ljStore_i,\ljExp \rangle \eval
  \langle \ljStore_i',\ljValv_i \rangle$
  then for all $\ljStore_j$, $\ljEnv_j$ with
  $\eqstore{(\ljStore_i,\ljEnv_i)}{(\ljStore_j,\ljEnv_j)}$
  $.$ $\ljEnv_j \entails \langle \ljStore_j,\ljExp \rangle \eval \langle
  \ljStore_j',\ljValv_j \rangle$ with
  $\eqstore{(\ljStore_i',\ljEnv_i)}{(\ljStore_j',\ljEnv_j)}$ and 
  $\eqstore{(\ljStore_i',\ljValv)}{(\ljStore_j',\ljValw)}$.
\end{lemma}

\begin{proof}
  Proof by induction on the derivation of $\ljExpe$.
\end{proof}

\subsection{Noninterference}

\begin{theorem}\label{thm:noninterference}
  For each $\ljEnv$ and $\ljStore$ with
  $\ljEnv \entails \langle \ljStore,\ljApp{(\ljFresh{\ljSbx\ljVar\ljExpe})}{\ljExpf}
  \rangle \eval \langle \ljStore',\ljVal\rangle$ it holds that $\eqenv\ljStore\ljStore'$.
\end{theorem}

\begin{proof}
  Proof by induction on the derivation of $\ljExpe$.
\end{proof}

\cleardoublepage
% _____      _       _           _  __          __        _    
%|  __ \    | |     | |         | | \ \        / /       | |   
%| |__) |___| | __ _| |_ ___  __| |  \ \  /\  / /__  _ __| | __
%|  _  // _ \ |/ _` | __/ _ \/ _` |   \ \/  \/ / _ \| '__| |/ /
%| | \ \  __/ | (_| | ||  __/ (_| |    \  /\  / (_) | |  |   < 
%|_|  \_\___|_|\__,_|\__\___|\__,_|     \/  \/ \___/|_|  |_|\_\

\section{Related work}
\label{sec:relatedwork}

There is a plethora of literature on securing JavaScript, so we focus on
the distinguishing features of our sandbox and on related work not
already discussed in the body of the paper.

\paragraph*{Sandboxing JavaScript}

The most closely related work to our sandbox mechanism is the design of access
control wrappers for revocable references and
membranes~\cite{CutsemMiller2010,Miller2006}. In a memory-safe language, a
function can only cause effects to objects reachable from references in parameters
and global variables. A revocable reference can be instructed to
detach from the objects, so that they are no longer reachable and safe
from effects. However, as membranes by themselves do not handle side effects (every
property access can be the call of a side-effecting getter) they do not
implement a sandbox in the way we did.

Agten et al.~\cite{AgtenVanAckerBrondsemaPhungDesmetPiessens2012} implement a
JavaScript sandbox using proxies and membranes. As in our work, 
they place wrappers around sensitive data (e.g., DOM elements) to enforce
policies and to prevent the application state from unprotected script inclusion.
However, instead of encapsulating untrusted code they require that
scripts are compliant with SES~\cite{SecureEcmaScript}, a subset of JavaScript's ``strict mode''
that prohibits features that are either unsafe or that grant uncontrolled
access, and use an SES-library to execute those scripts. A language-embedded
JavaScript parser transforms non-compliant scripts at run time, but
doing so restricts the handling of dynamic code compared to our approach. 

\TJS, a JavaScript contract system~\cite{KeilThiemann2015-treatjs},
uses a sandboxing mechanism similar to the 
sandbox presented in this work to guarantee that the execution of a predicate
does not interfere with the execution of a contract abiding host program. As in
our work, they use JavaScript's dynamic facilities to traverse the scope chain
when evaluating predicates and they use JavaScript proxies to make external
references visible when evaluating predicates. 

Arnaud et al.~\cite{ArnaudDenkerDucassePolletBergelSuen2010}
provide features similar to the sandboxing mechanism of
\TJS~\cite{KeilThiemann2015-treatjs}. Both approaches focus on access
restriction to guarantee side-effect
free contract assertion. However, neither of them implements a
full-blown sandbox, because writing is completely forbidden and always
leads to an exception.

Our sandbox works in a similar way and guarantees read-only access to target
objects, but redirects write operations to shadow objects such that local
modifications are only visible inside the sandbox. However, access restrictions
in all tree approaches affect only values that cross the border between two
execution environments. Values that are defined and used inside, e.g. local
values, were not restricted. Write access to those values is fine.

Patil et al. \cite{PatilDongLiLiangJiang2011} present \emph{JCShadow}, a
reference monitor implemented as a Firefox extension. Their tool provides
fine-grained access control to JavaScript resources. Similar to \SBX, they
implement shadow scopes that isolate scripts from each other and which regulate
the granularity of object access. Unlike \SBX, \emph{JCShadow} achieves a better
runtime performance. While more efficient, their approach
is platform-dependent as it is tied to a specific engine and requires
active maintenance to keep up with the development of the
enigine. \SBX, in contrast, is a JavaScript library based on 
the reflection API, which is part of the standard.

Most other approaches
(e.g., \cite{GoogleCaja,MillerSamuelLaurieAwadStay:caja_safe,FaecbookJS,ADsafe})
implement restrictions by filtering and rewriting untrusted code or by removing
features that are either unsafe of that grant uncontrolled access. For
exampe, Caja~\cite{GoogleCaja,MillerSamuelLaurieAwadStay:caja_safe}
compiles JavaScript code in a sanitized JavaScript subset that can safely be
executed on normal engines. Because static guarantees do not apply to code
created at run time using \lstinline{eval} or other mechanisms, Caja restricts
dynamic features and rewrites the code to a ``cajoled'' version with additional
run-time checks that prevent access to unsafe function and objects.

Static approaches come with a number of drawbacks, as shown
by a number of papers~\cite{MaffeisTaly2006,FeltHooimeijerEvansWeimer2008,PolitzEliopoulosGuhaKrishnamurthi2011}.
First, they either restrict the dynamic features of JavaScript or
their guarantees simply do not apply to
code generated at run time. Second, maintenance requires a
lot of effort because the implementation becomes obsolete as the language
evolves.

Thus, dynamic effect monitoring and dynamic access restriction plays an
important role in the context of JavaScript security, as shown
by a number of
authors~\cite{ArnaudDenkerDucassePolletBergelSuen2010,CutsemMiller2010,Miller2006,KeilThiemann2013-dls}.  

\paragraph*{Effect Monitoring}

Richards et al.~\cite{RichardsHammerNardelliJagannathan2013} provide a WebKit
implementation to monitor JavaScript programs at run time. Rather than performing
syntactic checks, they look at effects for history-based access control and 
revoke effects that violate policies implemented in C++.

Transcript, a Firefox extension by Dhawan et
al.~\cite{DhawanShanGanapathy2012}, extends JavaScript with support for
transactions and speculative DOM updates. Similar to \SBX, it builds a
transactional scope and permits the execution of unrestricted guest code. Effects
within a transaction are logged for inspection by the host program. They also
provide features to commit updates and to recover from effects of malicious
guest code.

JSConTest~\cite{HeideggerBieniusaThiemann2012-popl} is a framework that helps to
investigate the effects of unfamiliar JavaScript code by monitoring the
execution and by summarizing the observed access traces to access permission
contracts. It comes with an algorithm \cite{HeideggerThiemann2011} that infers a
concise effect description from a set of access paths and it enables the
programmer to specify the effects of a function using access permission
contracts.

JSConTest is implemented by an offline source code transformation. Because of
JavaScrip's flexibility it requires a lot of effort to construct an offline
transformation that guarantees full interposition and that covers the full
JavaScript language. This, the implementation of JSConTest has known omissions:
no support for \lstinline{with} and prototypes, and it does not apply to code
created at run time using \lstinline{eval} or other mechanisms.

JSConTest2 is a redesign and a reimplementation of
JSConTest using JavaScript proxies. The new implementation addresses
shortcomings of the previous version: it guarantees full interposition for the full language and for all
code regardless of its origin, including dynamically loaded code and code
injected via \lstinline{eval}.

JSConTest2 \cite{KeilThiemann2013-Proxy} monitors read and write operations on
objects through access permission contracts that specify allowed effects. A
contract restricts effects by defining a set of permitted access paths starting
from some anchor object. However, the approach works differently. JSConTest2 has
to encapsulate sensitive data instead of encapsulating dubious functions.

\paragraph*{Language-embedded Systems}

\textit{JSFlow}~\cite{HedinBirgissonBelloSabelfeld2014} is a full language-embedded
JavaScript interpreter that enforces information flow policies at run time. Like
\SBX, \textit{JSFlow} itself is implemented in JavaScript. Compared to
\SBX{}, the \textit{JSFlow} interpreter causes a significantly
higher run-time impact than the our sandbox, which only reimplements the
JavaScript semantics on the membrane.

A similar slowdown is reported for js.js~\cite{TerraceBeardKatta2012},
another language-embedded JavaScript interpreter conceived to execute untrusted
JavaScript code. Its implementation provides a wealth of powerful features
similar to \SBX{}: fine-grained access control, support for the full JavaScript
language, and full browser compatibility. However, its average
slowdown in the range of 100 to 200 is significantly higher than \SBX's.

\cleardoublepage
% _____                _   _           _   _____                 _ _       
%|  __ \              | | (_)         | | |  __ \               | | |      
%| |__) | __ __ _  ___| |_ _  ___ __ _| | | |__) |___  ___ _   _| | |_ ___ 
%|  ___/ '__/ _` |/ __| __| |/ __/ _` | | |  _  // _ \/ __| | | | | __/ __|
%| |   | | | (_| | (__| |_| | (_| (_| | | | | \ \  __/\__ \ |_| | | |_\__ \
%|_|   |_|  \__,_|\___|\__|_|\___\__,_|_| |_|  \_\___||___/\__,_|_|\__|___/

\section{Evaluation Results}
\label{sec:appendix/evaluation}

This section reports on our experience with applying the sandbox to select
programs. We focus on the influence of sandboxing on the execution time.

We use the Google Octane Benchmark Suite\footnote{\url{https://developers.google.com/octane}}
to measure the performance of the sandbox implementation. Octane measures a
JavaScript engine's performance by running a selection of complex and
demanding programs (benchmark programs run between 5 and 8200 times). 

Google claims that Octane ``measure[s] the performance of
JavaScript code found in large, real-world web applications, running on modern
mobile and desktop browsers.
Each benchmark is complex and demanding .

We use Octane as it is intended to measure the engine's performance (benchmark
programs run between 5 and 8200 times).
we claim that it is the heaviest kind of benchmark. 
Every real-world library (e.g.\ \lstinline{jQuery}) is less demanding and runs without an measurable runtime impact.

Octane 2.0 consists of 17 
programs\footnote{\url{https://developers.google.com/octane/benchmark}} that
range from performance tests to real-world web applications
(Figure~\ref{fig:evaluation/results}), from an OS kernel simulation to a
portable PDF viewer. Each program focuses on a special purpose, for example,
function and method calls, arithmetic and bit operations, array manipulation,
JavaScript parsing and compilation, etc.

\subsection{Testing Procedure}
\label{sec:evaluation/testingprocedure}

All benchmarks were run on a machine with two AMD Opteron processors with
2.20~GHz and 64~GB memory. All example runs and measurements reported in this
paper were obtained with the SpiderMonkey JavaScript engine.

For benchmarking, we wrote a new start script that loads and executes each
benchmark program in a fresh sandbox. By setting the sandbox global to the standard global
object, we ensure that each benchmark program can refer to properties
of the global object as needed.

As sandboxing wraps the global object in a membrane it mediates the
interaction of the benchmark program with the global application state.

All run time measurements were taken from a deterministic run, which requires
a predefined number of iterations\footnote{%
  Programs run either for one second or for a predefined number of iterations.
  If there are too few iterations in one second, it runs for another second.
}, and by using a warm-up run.

\subsection{Results}
\label{sec:evaluation/results}

\begin{figure}[t]
  \centering
  \small
  \begin{tabular}{l || r || r | r || r | r}
    \toprule
    \textbf{Benchmark}&
    \multicolumn{1}{c ||}{
      \textbf{Baseline}
    }&
    \multicolumn{2}{c ||}{%
      \textbf{Sandbox w/o Effects}
    }&
    \multicolumn{2}{c}{%
      \textbf{Sandbox w Effects}
    }\\

    \textit{}&
    \textit{time (sec)}&
    \textit{time (sec)}&
    \textit{slowdown}&
    \textit{time (sec)}&
    \textit{slowdown}\\
    \midrule

    Richards& 
    9 sec&
    12 sec&
    1.33&
    15 sec&
    1.67\\

    DeltaBlue&
    9 sec&
    -&
    -&
    -&
    -\\

    Crypto&
    18 sec&
    42 sec&
    2.33&
    88 sec&
    4.89\\

    RayTrace&
    9 sec&
    74 sec&
    8.22 &
    498 sec&
    55.33\\

    EarleyBoyer&
    19 sec&
    202 sec&
    10.63&
    249 sec&
    13.11\\
    
    RegExp&
    6 sec&
    9 sec&
    1.5&
    12 sec&
    2\\

    Splay&
    3 sec&
    19 sec&
    6.33&
    33 sec&
    11\\
    
    SplayLatency&
    3 sec&
    19 sec&
    6.33&
    33 sec&
    11\\

    NavierStokes&
    3 sec&
    56 sec&
    18.67&
    61 sec&
    20.33\\

    pdf.js&
    7 sec&
    113 sec&
    16.14&
    778 sec&
    111.14\\

    Mandreel&
    8 sec&
    151 sec&
    18.88&
    483 sec&
    60.38\\

    MandreelLatency&
    8 sec&
    151 sec&
    18.88&
    483 sec&
    60.38\\

    Gameboy Emulator&
    4 sec&
    17 sec&
    4.25&
    26 sec&
    6.50\\

    Code loading &
    8 sec&
    11 sec&
    1.38&
    12 sec&
    1.50\\

    Box2DWeb&
    4 sec&
    145 sec&
    36.25&
    1,302 sec&
    325.50\\
    
    zlib&
    7 sec&
    -&
    -&
    -&
    -\\

    TypeScript&
    26 sec&
    61 sec&
    2.35&
    328 sec&
    12.62\\
    
    \midrule
    Total&
    135 sec&
    1.082 sec&
    8.01&
    4,401 sec&
    32.60\\

    \bottomrule
  \end{tabular}
  \caption[Timings from running the Google Octane 2.0 Benchmark Suite.]{Timings from running the Google Octane 2.0 Benchmark Suite. The first column \textbf{Baseline} gives the baseline execution times without sandboxing. The column \textbf{Sandbox w/o Effects} shows the time required to complete a sandbox run without effect logging and the relative slowdown (Sandbox time/Baseline time). The column \textbf{Sandbox w Effects} shows the time and slowdown (w.r.t.\ Baseline) of a run with fine-grained effect logging.}
  \label{fig:DecentJS/Evaluation/Results/Timings}
\end{figure}

\begin{figure}[t]
  \centering
  \small
  \begin{tabular}{l || r || r || r | r | r}
    \toprule
    \textbf{Benchmark}&
    \textbf{Objects}&
    \textbf{Effects}&
    \multicolumn{3}{c}{%
      \textbf{Size of Effect List}
    }\\
    \textit{}&
    \textit{}&
    \textit{}&
    \textit{Reads}&
    \textit{Writes}&
    \textit{Calls}\\
    \midrule

    Richards& 
    14&
    492073&
    20&
    2&
    5\\

    DeltaBlue&
    -&
    -&
    -&
    -&
    -\\

    Crypto&
    21&
    4964248&
    29&
    2&
    11\\

    RayTrace&
    18&
    51043282&
    26&
    3&
    8\\

    EarleyBoyer&
    33&
    4740377&
    42&
    8&
    6\\
    
    RegExp&
    16&
    296995&
    23&
    2&
    6\\

    Splay&
    16&
    1635732&
    23&
    2&
    8\\
    
    SplayLatency& 
    16&
    1635732&
    23&
    2&
    8\\

    NavierStokes&
    15&
    4089&
    21&
    2&
    6\\

    pdf.js&
    36&
    77665629&
    59&
    8&
    21\\

    Mandreel&
    31&
    39948598&
    50&
    2&
    21\\

    MandreelLatency& 
    31&
    39948598&
    50&
    2&
    21\\

    Gameboy Emulator&
    28&
    1225935&
    42&
    2&
    16\\

    Code loading &
    12417&
    107481&
    50&
    2&
    13\\

    Box2DWeb&
    28&
    132722198&
    38&
    2&
    14\\
    
    zlib&
    -&
    -&
    -&
    -&
    -\\

    TypeScript&
    23&
    27518481&
    34&
    2&
    9\\
    
    \midrule
    Total&
    12743&
    383949448&
    530&
    43&
    173\\

    \bottomrule
  \end{tabular}
  \caption[Numbers from internal counters.]{Numbers from internal counters. Column \textbf{Objects} shows the numbers of wrap objects and column \textbf{Effects} gives the total numbers of effects. Column \textbf{Size of Effect List} lists the numbers of \emph{different} effects after running the benchmark. Column \textbf{Reads} shows the number of read effects distinguished from there number of write effects (Column \textbf{Writes}) and distinguished from there number of call effects (Column \textbf{Calls}). Multiple effects to the same field of an object are counted as one effect.}
    \label{fig:DecentJS/Evaluation/Results/Statisics}
\end{figure}

Figure~\ref{fig:DecentJS/Evaluation/Results/Timings} and Figure~\ref{fig:DecentJS/Evaluation/Results/Statisics} 
contains the run-time statistics for all
benchmark programs in two different configurations, which are explained in the
figure's caption, and lists the readouts of some internal counters.
Multiple read effects to the same field of an object are counted as one effect.

As expected, the run time increases when executing a benchmark in a sandbox. While
some programs like \emph{EarleyBoyer}, \emph{NavierStrokes}, \emph{pdf.js},
\emph{Mandreel}, and \emph{Box2DWeb} are heavily affected, others are
only slightly
affected: \emph{Richards}, \emph{Crypto}, \emph{RegExp}, and \emph{Code
loading}, for instance.
Unfortunately, \emph{DeltaBlue} and \emph{zlib} do not run in our sandbox.
\emph{DeltaBlue} attempts to add a new property to the global
\mbox{\lstinline{Object.prototype}} object. As our sandbox prevents
unintended modifications to
\mbox{\lstinline{Object.prototype}} the new property is
only visible inside of the current sandbox and only to objects created with
\mbox{\lstinline{new Object()}} and \mbox{\lstinline{Object.create()}}, but not to those
created using object literals.  

The \emph{zlib} benchmark uses an indirect call\footnote{%
  An indirect call invokes the \lstinline{eval} function by using a name other
  than \lstinline{eval}.
} to \lstinline{eval} to write objects to the global scope, which is not allowed
by the ECMAScript 6 (ECMA-262) specification. Another benchmark, \emph{Code
loading}, also uses an indirect call to \lstinline{eval}. A small
modification makes the program compatible with the normal \lstinline{eval},
which can safely be used in our sandbox.

In the first experiment we turn off effect logging, whereas in the second one it
remains enabled. Doing so separates the performance impact of the sandbox system
(proxies and shadow objects) from the impact caused by the effect system. From
the running times we find that the sandbox itself causes an average slowdown of
8.01 (over all benchmarks).

Our experimental setup wraps the global object in a membrane
and mediates the interaction between the benchmark program and the global
application state. As each benchmark program contains every source required to
run the benchmark in separation, except global objects and global functions,
the only thing that influences the execution time is read/write access to global
elements.

In absolute times, raw sandboxing causes a run time deterioration
of 0.003ms per sandbox operation (effects) (0.011ms with effect logging
enabled). For example, the \emph{Box2DWeb} benchmark requires 145 seconds to
complete and performs 132,722,198 effects on its membrane. Its baseline requires 4
seconds. Thus, sandboxing takes an additional 141 seconds. Hence,
there is an overhead of 0.001ms per operation
(0.010ms with effect logging enabled).

The results from the tests also indicate that the garbage collector runs more
frequently, but there is no significant increase in the memory consumption.
For the effect-heaviest benchmark \emph{Box2D} we find that the \emph{virtual memory
size} increases from 157MByte (raw run) to 197MB (full run with effect logging). 

However, when looking at all benchmarks, the difference in the virtual memory
size compared with the baseline run ranges from -126MByte to +40MByte for a raw
sandbox run without effect logging and from -311MByte to +158MByte for a full
run with fine grained effect logging. Appendix~\ref{sec:appendix/results/memory}
shows the memory usage of the different benchmark programs and their difference
compared with the baseline.

\subsection{Google Octane Scores Values}
\label{sec:appendix/results/score}

Octane reports its result in terms of a score. The Octane FAQ\footnote{\url{https://developers.google.com/octane/faq}} explains the score as follows: ``\emph{In a nutshell: bigger is better. Octane measures the time a test takes to complete and then assigns a score that is inversely proportional to the run time.}'' The constants in this computation are chosen so that the current overall score (i.e., the geometric mean of the individual scores) matches the overall score from earlier releases of Octane and new benchmarks are integrated by choosing the constants so that the geometric mean remains the same. The rationale is to maintain comparability.

\begin{figure}[t]
  \centering
  \small
  \begin{tabular}{ l || r | r || r}
    \toprule
    \textbf{Benchmark}& 
    \textbf{Isolation}&
    \textbf{Effects}&
    \textbf{Baseline}\\
    \midrule
    Richards&
    4825&
    4135&
    6552\\
    
    DeltaBlue&
    -&
    -&
    6982\\

    Crypto&
    2418&
    1131&
    5669\\

    RayTrace&
    1387&
    179&
    9692\\

    EarleyBoyer&
    954&
    767&
    10345\\

    RegExp&
    911&
    1139&
    1535\\

    Splay&
    1268&
    713&
    8676\\

    SplayLatency&
    3630&
    1818&
    12788\\

    NavierStokes&
    989&
    890&
    15713\\

    pdf.js&
    434&
    63&
    8182\\

    Mandreel&
    346&
    106&
    9102\\

    MandreelLatency&
    2518&
    526&
    12526\\

    Gameboy Emulator&
    6572&
    4780&
    31865\\

    Code loading &
    7348&
    6000&
    9136\\

    Box2DWeb&
    453&
    50.1&
    18799\\

    zlib&
    -&
    -&
    42543\\

    TypeScript&
    4554&
    792&
    12588\\
    
    \bottomrule
  \end{tabular}
  \caption{%
    Scores for the Google Octane 2.0 Benchmark Suite (bigger is better). Block \textbf{Isolation} contains the score values of a raw sandbox run without effect logging, whereas block \textbf{Effects} contains the score values of a full run with fine-grained effect logging. The last column \textbf{Baseline} gives the baseline scores without sandboxing.}
  \label{fig:appendix/benchmarks/scores}
\end{figure}

Figure~\ref{fig:appendix/benchmarks/scores} contains the scores of all benchmark programs in different configurations, which are explained in the figure's caption. All scores were taken from a deterministic run, which requires a predefined number of iterations\footnote{%
  Programs run either for one second or for a predefined number of iterations. If there are to few iterations in one second, it runs for another second.}, and by using a warm-up run.

As expected, all scores drop when executing the benchmark in a sandbox. In the first experiment, we turn off effect logging, whereas the second run is with effect logging. This splits the performance impact into the impact caused by the sandbox system (proxies and shadow objects) and the impact caused by effect system.

\subsection{Memory Consumption}
\label{sec:appendix/results/memory}

\begin{figure}[t]
  \centering
  \small
  \begin{tabular}{ l || r || r || r || r}
    \toprule
    \textbf{Benchmark}& 
    \multicolumn{4}{c}{
      \textbf{Baseline}
    }\\

    \textbf{}&

    % Baseline
    \multicolumn{1}{c ||}{
      \textbf{Virtual}
    }&
    \multicolumn{1}{c ||}{
      \textbf{Resident}
    }&
    \multicolumn{1}{c ||}{
      \textbf{Text/Data}
    }&
    \multicolumn{1}{c}{
      \textbf{Shared}
    }\\

    \textbf{}&
    \textit{size}&
    \textit{size}&
    \textit{size}&
    \textit{size}\\
    \midrule

    Richards&
    134&	19&	109&	5\\

    DeltaBlue&
    -&	-&	-&	-\\

    Crypto&
    225&	105&	201&	6\\

    RayTrace&
    148&	31&	124&	5\\

    EarleyBoyer&
    500&	363&	476&	6\\

    RegExp&
    226&	108&	202&	6\\

    Splay&
    535&	416&	511&	6\\

    SplayLatency&
    535&	416&	511&	6\\

    NavierStokes&
    141&	24&	116&	5\\

    pdf.js&
    316&	169&	292&	6\\

    Mandreel&
    305&	182&	280&	6\\

    MandreelLatency&
    305&	182&	280&	6\\

    Gameboy Emulator&
    194&	62&	170&	6\\

    Code loading &
    268&	142&	243&	6\\

    Box2DWeb&
    157&	53&	132&	5\\

    zlib&
    -&	-&	-&	-\\

    TypeScript&
    473&	369&	448&	6\\

    \bottomrule
  \end{tabular}
  \caption[Memory usage when running the Google Octane 2.0 Benchmark Suite.]{Memory usage when running the Google Octane 2.0 Benchmark Suite without sandboxing. Column \textbf{Virtual} shows the virtual memory size, column \textbf{Resident} shows the resident set size, column \textbf{Text/Data} shows the Text/Data segment size, and column \textbf{Text/Data} shows the Text/Data segment size. All values are in MByte.}
  \label{fig:DecentJS/Evaluation/Memory/Baseline}
\end{figure}

\begin{figure}[t]
  \centering
  \small
  \begin{tabular}{ l || r | r || r | r || r | r || r | r}
    \toprule
    \textbf{Benchmark}& 
    \multicolumn{8}{c}{%
      \textbf{Sandbox w/o Effects}
    }\\

    \textbf{}&

    % Sandbox w/o Effects
    \multicolumn{2}{c ||}{
      \textbf{Virtual}
    }&
    \multicolumn{2}{c ||}{
      \textbf{Resident}
    }&
    \multicolumn{2}{c ||}{
      \textbf{Text/Data}
    }&
    \multicolumn{2}{c}{
      \textbf{Shared}
    }\\

    \textbf{}&

    \textit{size}&
    \textit{diff.}&
    \textit{size}&
    \textit{diff.}&
    \textit{size}&
    \textit{diff.}&
    \textit{size}&
    \textit{diff.}\\
    \midrule

    Richards&
    135&	1&	20&	1&	110&	1&	5&	0\\

    DeltaBlue&
    -&	-&	-&	-&	-&	-&	-&	-\\

    Crypto&
    232&	+7&	106&	+1&	208&	+7&	6&	0\\

    RayTrace&
    148&	+0&	31&	+0&	124&	+0&	6&	+1\\

    EarleyBoyer&
    374&	-126&	272&	-91&	350&	-126&	6&	0\\

    RegExp&
    224&	-2&	107&	-1&	200&	-2&	6&	0\\

    Splay&
    466&	-69&	352&	-64&	422&	-89&	6&	0\\

    SplayLatency&
    466&	-69&	352&	-64&	422&	-89&	6&	0\\

    NavierStokes&
    134&	-7&	18&	-6&	109&	-7&	5&	0\\

    pdf.js&
    274&	-42&	123&	-46&	250&	-42&	6&	0\\

    Mandreel&
    263&	-42&	133&	-49&	239&	-41&	5&	-1\\

    MandreelLatency&
    263&	-42&	133&	-49&	239&	-41&	5&	-1\\

    Gameboy Emulator&
    188&	-6&	60&	-2&	163&	-7&	6&	0\\

    Code loading&
    259&	-9&	138&	-4&	234&	-9&	6&	0\\

    Box2DWeb&
    197&	+40&	97&	+44&	172&	+40&	6&	+1\\

    zlib&
    -&	-&	-&	-&	-&	-&	-&	-\\

    TypeScript&
    424&	-49&	325&	-44&	428&	-20&	6&	0\\

    \bottomrule
  \end{tabular}
  \caption[Memory usage when running the Google Octane 2.0 Benchmark Suite.]{%
    Memory usage of a raw sandbox run without effect logging. Column \textbf{Virtual} shows the virtual memory size, column \textbf{Resident} shows the resident set size, column \textbf{Text/Data} shows the Text/Data segment size, and column \textbf{Text/Data} shows the Text/Data segment size. Sub-column \textit{size} shows the size in MByte and sub-column \textit{diff.} shows the difference to the baseline (Sandbox size - Baseline size) in MByte.}
  \label{fig:DecentJS/Evaluation/Memory/SandboxWithoutEffects}
\end{figure}

\begin{figure}[t]
  \centering
  \small
  \begin{tabular}{ l || r | r || r | r || r | r || r | r}
    \toprule
    \textbf{Benchmark}& 
    \multicolumn{8}{c}{%
      \textbf{Sandbox w Effects}
    }\\

    \textbf{}&

    % Sandbox w Effects
    \multicolumn{2}{c ||}{
      \textbf{Virtual}
    }&
    \multicolumn{2}{c ||}{
      \textbf{Resident}
    }&
    \multicolumn{2}{c ||}{
      \textbf{Text/Data}
    }&
    \multicolumn{2}{c}{
      \textbf{Shared}
    }\\

    \textbf{}&

    \textit{size}&
    \textit{diff.}&
    \textit{size}&
    \textit{diff.}&
    \textit{size}&
    \textit{diff.}&
    \textit{size}&
    \textit{diff.}\\
    \midrule

    Richards&
    167&	33&	54&	35&	142&	33&	6&	1\\

    DeltaBlue&
    -&	-&	-&	-&	-&	-&	-&	-\\

    Crypto&
    209&	-16&	93&	-12&	184&	-17&	6&	0\\

    RayTrace&
    181&	+33&	63&	+32&	157&	+33&	6&	+1\\

    EarleyBoyer&
    189&	-311&	86&	-277&	195&	-281&	6&	0\\

    RegExp&
    224&	-2&	104&	-4&	200&	-2&	6&	0\\

    Splay&
    424&	-111&	321&	-95&	399&	-112&	6&	0\\

    SplayLatency&
    424&	-111&	321&	-95&	399&	-112&	6&	0\\

    NavierStokes&
    134&	-7&	19&	-5&	109&	-7&	5&	0\\

    pdf.js&
    272&	-44&	116&	-53&	248&	-44&	6&	0\\

    Mandreel&
    347&	+42&	160&	-22&	323&	+43&	6&	0\\

    MandreelLatency&
    347&	+42&	160&	-22&	323&	+43&	6&	0\\

    Gameboy Emulator&
    214&	+20&	84&	+22&	190&	+20&	6&	0\\

    Code loading &
    262&	-6&	139&	-3&	238&	-5&	6&	0\\

    Box2DWeb&
    191&	+34&	72&	+19&	166&	+34&	6&	+1\\

    zlib&
    -&	-&	-&	-&	-&	-&	-&	-\\

    TypeScript&
    631&	+158&	493&	+124&	607&	+159&	6&	0\\

    \bottomrule
  \end{tabular}
  \caption[Memory usage when running the Google Octane 2.0 Benchmark Suite.]{%
    Memory usage of a full run with fine-grained effect logging. Column \textbf{Virtual} shows the virtual memory size, column \textbf{Resident} shows the resident set size, column \textbf{Text/Data} shows the Text/Data segment size, and column \textbf{Text/Data} shows the Text/Data segment size. Sub-column \textit{size} shows the size in MByte and sub-column \textit{diff.} shows the difference to the baseline (Sandbox size - Baseline size) in MByte.}
  \label{fig:DecentJS/Evaluation/Memory/SandboxWithEffects}
\end{figure}

Figure~\ref{fig:DecentJS/Evaluation/Memory/Baseline}, Figure~\ref{fig:DecentJS/Evaluation/Memory/SandboxWithoutEffects}, and Figure~\ref{fig:DecentJS/Evaluation/Memory/SandboxWithEffects} shows the memory consumption recorded when running the Google Octane 2.0 Benchmark Suite. The numbers indicate that there is no significant increase in the memory consumed. For example, the difference of the virtual memory size ranges from -126 to 40 for a raw sandbox run and from -311 to +158 for a full run with fine grained effect logging.

%%
%% Draft Material
%%

%\input{draft}

\end{document}